\documentclass[journal]{IEEEtran}
\usepackage{float}
\usepackage{graphicx}
\usepackage{textcomp}
\usepackage{colortbl}
\usepackage{adjustbox}
\usepackage{caption}
\usepackage{paralist}
\usepackage{tabularx}
\usepackage{import}
\usepackage{algpseudocode,algorithm}
\usepackage{listings}
\usepackage{multirow}
\usepackage{caption}
\usepackage{subcaption}
\usepackage{hyperref}
\usepackage{circuitikz}
\usepackage{pgf}
\usepackage{pgfplots}
\usepackage{xcolor}
\usepackage{bm}
\usepackage{pgfplotstable}
\usepackage[many]{tcolorbox}
\pgfplotsset{compat=1.8}
\usepackage{adjustbox}
\usepackage{changepage}
\usepackage{enumitem}
\usepackage{dblfloatfix}
\usepackage{minted}
\usepackage{svg}
\usepackage{xpatch}

\usepackage[utf8]{inputenc}
\usepackage[T1]{fontenc}

\usepackage{booktabs}

\tcbset{
  highlightbox/.style={
    colback=gray!10,              % fill colour
    colframe=gray!50,             % border colour
    boxrule=0.4pt,                % border width
    arc=5pt,                      % corner radius
    left=5pt,right=5pt,top=3pt,bottom=3pt, % padding
    enhanced,                     % enables rounded corners
  }
}

\ifCLASSOPTIONcompsoc
  \usepackage[nocompress]{cite}
\else
  \usepackage{cite}
\fi

\ifCLASSINFOpdf
\else
\fi
\begin{document}
\title{Power Side-Channel Membership Inference Attack on Embedded Machine Learning}

\author{Sahan~Sanjaya,~\IEEEmembership{Student Member,~IEEE} and~Prabhat~Mishra,~\IEEEmembership{Fellow,~IEEE \vspace{-0.1in}}

% \IEEEcompsocitemizethanks{\IEEEcompsocthanksitem S. Sanjaya, A. Jayasena and P. Mishra are with the Department of Computer \& Information Science \& Engineering, University of Florida, USA. Email: \{ssanjaya, arunajayasena, prabhat\}@ufl.edu}
% }

\IEEEcompsocitemizethanks{\IEEEcompsocthanksitem S. Sanjaya and P. Mishra are with the Department of Computer \& Information Science \& Engineering, University of Florida, USA.%
\ Email: \{ssanjaya, prabhat\}@ufl.edu.}
}



\maketitle

\begin{abstract}
Membership inference attacks (MIAs) threaten the privacy of machine learning (ML) training data by determining whether a sample was used to train a target model. Existing MIAs rely on model outputs, ranging from prediction probabilities to predicted labels, an assumption that can be restrictive for on-device ML systems with limited or inaccessible outputs. However, suppressing model outputs does not eliminate the data-dependent computations that produce them, which may remain observable through physical side channels.

We present \emph{PSCMIA}, a power side-channel membership inference attack against embedded ML models that can infer membership directly from power traces without requiring prediction probabilities or even the predicted labels. We evaluate PSCMIA across multiple datasets (MNIST, FMNIST, CIFAR10, CINIC10), fully connected (FC) and convolutional neural network (CNN) architectures, and two embedded platforms (STM32F3, XMEGA). PSCMIA achieves ROC-AUC values of up to $0.907$ on FC models. For CNN models, the ROC-AUC gap between PSCMIA and probability vector-based shadow MIA ranges from $0.006$ to $0.116$. Across the FC and CNN evaluations, PSCMIA outperforms label-only MIA in 11 of 16 model-dataset-hardware configurations, demonstrating that physical execution can expose membership information even when conventional model outputs are unavailable through unintended power side-channel leakage.
\end{abstract}


% \fi
\IEEEpeerreviewmaketitle

\pagestyle{plain}

\section{Introduction}\label{sec:intro}

The rapid advancement of machine learning (ML) has enabled its adoption across a wide range of applications, including privacy-sensitive domains, such as healthcare, finance, and defense. Developing an ML model requires a training dataset, which may contain sensitive or confidential information. Although the training data are not exposed to users during inference, determining whether a particular sample was included in the training set can itself reveal sensitive information. This privacy risk has motivated the development of membership inference attacks (MIAs), which aim to determine whether a given sample was used to train a target model~\cite{shokri2017membership}.

MIAs exploit the observation that an ML model may exhibit distinguishable behavior for training samples compared with previously unseen samples. Early MIAs commonly employed shadow models that approximate the behavior of the target model in an adversary-controlled environment, allowing the adversary to learn membership-dependent characteristics from model outputs. Subsequent work has demonstrated that membership information can be inferred from several output-derived properties, including the prediction probability vector, prediction confidence, entropy, prediction correctness, and distance to the decision boundary. As defenses have increasingly restricted access to detailed model outputs, MIA techniques have evolved from attacks requiring the complete prediction probability vector to more restrictive settings in which only the predicted label is available.

On-device inference is popular in embedded systems where ML models execute locally rather than relying on cloud-based inference services. Local execution reduces communication overhead and enables ML deployment in latency-sensitive and intermittently connected applications. However, deploying ML models on embedded devices introduces stringent constraints on memory, computation, and energy consumption. Embedded ML systems and resource-efficient inference frameworks such as TinyML~\cite{warden2019tinyml} and BitNetMCU~\cite{bitnetmcu} address these constraints through compact model architectures, quantization, and optimized embedded implementations. 
%Consequently, increasingly capable ML models can now execute directly on resource-constrained devices, including wearable, sensing, and other privacy-sensitive embedded platforms.

Embedded devices are well known to leak information through physical side-channels, including power consumption, electromagnetic emanations, timing, and acoustic signals. Although side-channel attacks were historically developed primarily to recover cryptographic secrets, recent studies have demonstrated that physical leakage can also reveal non-cryptographic information, including executed instructions, user activity, program behavior, and properties of ML computations~\cite{sanjaya2025application}. In particular, power side-channel analysis of embedded ML implementations has been shown to expose information about model architectures, implementation parameters, and processed inputs. These observations raise an important question: \textbf{\textit{Can the power side-channel generated during embedded ML inference reveal whether an input was part of the model’s training set?}}

This question is particularly relevant for embedded ML systems because the assumptions underlying conventional output-based MIAs may not hold. An adversary targeting a remote ML service may obtain prediction probabilities, confidence scores, or at least the predicted label through an exposed inference interface. In contrast, an embedded ML model may expose only a restricted application-level output, or may not externally reveal the model prediction at all. Consequently, directly applying conventional output-based MIAs can be difficult or impossible in such settings.

% \color{blue}
% The attack proposed in~\cite{ehsan2026side} uses software-accessible power measurements on mobile devices to infer membership in  ML agents by exploiting power variations arising from mechanisms such as early exit, pruning, and dynamic routing. However, their threat model assumes that the adversary can inject crafted samples into a shared training corpus and later run an unprivileged co-resident application to obtain energy measurements through standard Android kernel interfaces. \textcolor{red}{This assumption limits the applicability of this attack ....}
% \color{black}

Restricting access to the model output does not eliminate the underlying information exploited by MIAs. The prediction probability and other output-derived properties are not independent artifacts generated only at the inference interface. Rather, they arise from a sequence of data-dependent computations that modulate the instantaneous power consumption of the processor. \textit{We hypothesize that power traces collected during inference can retain membership-dependent information correlated with the internal computations that  determine the model output, even when detailed model outputs are unavailable to the adversary.}

In this paper, we investigate this hypothesis by introducing a \emph{power side-channel membership inference attack (\textbf{PSCMIA})} against embedded ML models. Instead of relying on prediction probabilities or confidence scores exposed by the model interface, the proposed attack extracts membership information directly from power traces collected during inference. Our results show that the power side-channel of an execution of an embedded ML model contains sufficient information to distinguish training members from non-members, establishing physically measured power traces as a new membership-inference channel for embedded ML systems.

%\subsection{Contributions}\label{subsec:contribution}

\textbf{Contributions.} The main contributions are as follows:

\begin{compactitem} 
\setlength{\itemsep}{6pt}

\item We introduce PSCMIA, a power side-channel membership inference attack for embedded ML that infers membership without access to prediction probabilities, confidence scores, or even the predicted label. 

\item We evaluate PSCMIA across multiple datasets, two model architectures, and two hardware platforms, and compare it with probability vector-based shadow MIA and label-only MIA. PSCMIA achieves ROC-AUC values of up to $0.9065$, outperforming label-only MIA in most evaluated settings while approaching shadow MIA performance in several cases. 

\item We show that PSCMIA effectiveness correlates with target-model vulnerability to probability vector-based shadow MIA, with highly vulnerable models exhibiting stronger power side-channel membership leakage.

\item We characterize membership-relevant leakage in inference-time power traces, showing that the traces expose information correlated with predicted class, correctness, confidence, and margin. We further localize stronger member/non-member separation near the final decision computations.

\item We develop an end-to-end profiling-based attack framework that combines shadow-model profiling, leakage localization, feature extraction, and membership classification using power traces. 

\item We evaluate the impact of varying levels of target-model overfitting and prediction correctness on PSCMIA and attack model transferability across datasets.

\end{compactitem}

To the best of our knowledge, \textit{\textbf{this work presents the first MIA against embedded ML models that relies solely on physically measured power traces, without requiring access to any target-model outputs.}} The remainder of this paper is organized as follows. Section~\ref{sec:related} provides relevant background and surveys related efforts. Section~\ref{sec:methodology} describes the proposed attack methodology. Section~\ref{sec:experiments} outlines the experimental setup followed by the experimental results in Section~\ref{sec:results}. Section~\ref{sec:limits} discusses several limitations of the proposed attack. Finally, Section~\ref{sec:conclusion} concludes the paper.

\section{Background and Related Work}
\label{sec:related}

In this section, we first provide background on MIAs, power side-channel analysis, and embedded ML. Next, we survey related efforts to highlight our contributions.

\vspace{-0.15in}
\subsection{Membership Inference Attacks}\label{subsec:related_mia}

Membership inference attacks (MIAs) determine whether a candidate sample was included in the training data of a target model by exploiting behavioral differences between members and previously unseen inputs~\cite{hu2022membership}. Early black-box attacks used shadow models to learn these differences from prediction probability vectors~\cite{shokri2017membership}, while subsequent approaches used simpler signals such as prediction confidence~\cite{salem2018ml}, loss~\cite{yeom2018privacy}, entropy~\cite{salem2018ml}, correctness~\cite{yeom2018privacy}, and the model's sensitivity to input perturbations~\cite{choquette2021label}. Recent work has expanded the threat model in several directions, including white-box attacks that access gradients and intermediate model computations~\cite{nasr2019comprehensive}, label-only attacks that estimate distance to the decision boundary~\cite{li2021membership, choquette2021label, li2025enhanced}, and likelihood-ratio attacks that compare the target sample against reference models~\cite{carlini2022membership}. MIAs have also been extended to generative~\cite{hayes2017logan} and large language models through neighborhood comparisons~\cite{mattern2023membership}, and token-level behavior~\cite{shi2024detecting}. Overall, the literature has progressively reduced the information required from the target model while improving attack reliability.

\vspace{-0.15in}
\subsection{Power Side-Channel Attacks}\label{subsec:related_psc}

Power side-channel attacks exploit the relationship between a device's power consumption and the data or operations processed during execution. Classical techniques, including simple power analysis~\cite{kocher1999differential}, differential power analysis~\cite{kocher1999differential}, and correlation power analysis~\cite{brier2004correlation}, commonly rely on statistical relationships between measured traces and hypothesized intermediate values. More recent profiled attacks instead train statistical or ML models on labeled traces collected from a similar device and subsequently apply the learned leakage profile to a target device \cite{heuser2012intelligent, lerman2015machine, gilmore2015neural, maghrebi2016breaking}.

Power side-channel research has focused mainly on extracting secret keys from cryptographic implementations~\cite{sanjaya2024information, sanjaya2025sleepwalk, sanjaya2026sleep}. However, the attack surface has expanded to several application domains~\cite{sanjaya2025application}, including user-activity fingerprinting~\cite{chen2017powerful, taneja2023hot}, instruction-level disassembly~\cite{park2018power, glamovcanin2023instruction }, and ML systems \cite{xiang2020open, gao2024deeptheft}. Existing attacks against ML implementations primarily target model confidentiality and input privacy. Software-based attacks use interfaces such as Intel RAPL and NVIDIA's power-monitoring utilities to recover model architectures, layer types, and hyperparameters \cite{gao2024deeptheft,ryu2023gamma,jha2020deeppeep}. Remote attacks on FPGA accelerators employ on-chip voltage sensors to identify layer boundaries, folding factors, neuron counts, and accelerator operations \cite{zhang2021stealing,meyers2023active,tian2023practical}. Physical attacks have similarly recovered model types, architectural parameters, model weights, and private inference inputs from CPUs, MCUs, FPGAs, ASICs, and compute-in-memory accelerators \cite{xiang2020open,wei2018know,wolf2021stealing,wang2023powergan}. Collectively, these studies demonstrate that ML execution produces distinguishable power signatures that encode both model-dependent and input-dependent information. Nevertheless, their primary objectives are model extraction or input reconstruction rather than determining whether a candidate sample was used to train the model.

\vspace{-0.15in}
\subsection{Embedded Machine Learning}\label{subsec:embed_ml}

Embedded machine learning executes trained models directly on resource-constrained edge platforms under strict memory, latency, and energy budgets. It enables low-latency and often offline applications, including anomaly detection, wearable sensing, smart-city monitoring, and automotive and medical systems~\cite{sanchez2020tinyml, banbury2021mlperf,capogrosso2024machine,soro2021tinyml, bitnetmcu}. Although on-device inference reduces dependence on cloud services, it does not inherently guarantee privacy. Embedded devices frequently process sensitive sensor data and may be physically or locally accessible to adversaries, while their data-dependent execution can expose information about model inputs, prediction behavior, and training-set membership~\cite{sanjaya2025application}. Such leakage can have serious privacy consequences because determining that a record was used during training may reveal an individual's participation in sensitive healthcare, location, or behavioral datasets, even when the model parameters and prediction probability vectors remain hidden.

\begin{figure*}[t]
    \centering
    \begin{subfigure}{0.98\textwidth}
        \centering
        \includegraphics[width=\textwidth]{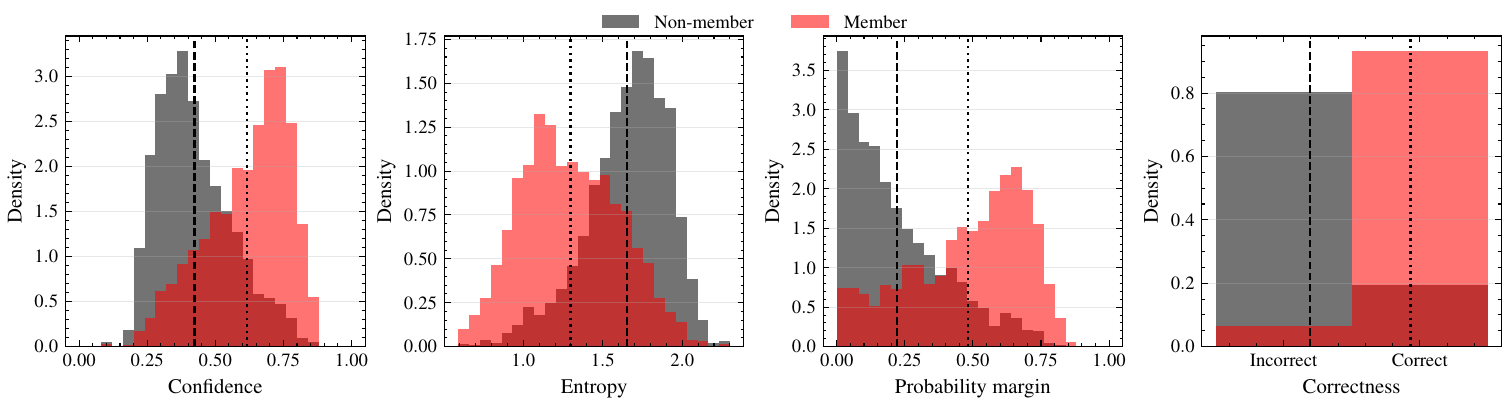}
        \caption{Software-computed probability vector properties}
        \label{fig:true_property_member_nonmember_hist}
    \end{subfigure}

    \begin{subfigure}{0.98\textwidth}
        \centering
        \includegraphics[width=\textwidth]{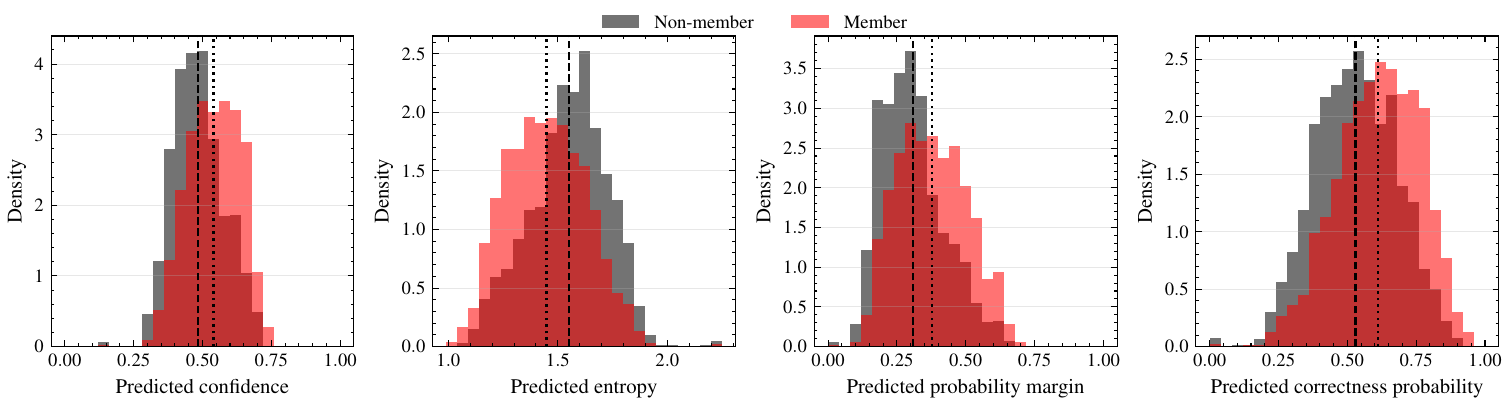}
        \caption{Trace-predicted probability vector properties}
        \label{fig:trace_pred_property_member_nonmember_hist}
    \end{subfigure}

    \caption{Member and non-member distributions for true and trace-predicted
    output properties. The top panel shows that the selected model exhibits
    membership-dependent differences in software-computed properties. The bottom panel shows that
    power-trace-based probes preserve part of this separation, supporting the
    hypothesis that power traces leak information correlated with the
    output-probability distribution-derived signals used by output-based membership inference attacks.}
    \label{fig:motivation_member_nonmember_property_histograms}
\end{figure*}

\vspace{-0.15in}
\subsection{Related Work}\label{subsec:related}

Few studies have investigated membership leakage through side channels rather than intended model outputs. Timing-based attacks exploit data-dependent execution latency~\cite{shukla2023whispering}, while {GateBleed}~\cite{kalyanapu2025gatebleed} uses latency caused by accelerator power gating, and system-level privacy side channels exploit input processing, filtering, or monitoring components. {GhostTrace}~\cite{srivastava2026ghosttrace} extends this direction by using GPU hardware performance counters to classify members without accessing model outputs or internal parameters. A power-based MIA against energy-efficient mobile agents was presented in~\cite{ehsan2026side}, where early exits, adaptive pruning, and dynamic routing produce membership-dependent energy patterns observable through Android software interfaces. The attack assumes that the adversary can contribute crafted samples to a shared training corpus before model training and can later run an unprivileged co-resident application on the target device to collect energy measurements through standard Android kernel interfaces. These studies establish execution behavior as an independent source of membership leakage.

Existing side-channel-based MIAs have not investigated externally measured power traces from embedded ML systems. Our work fills this gap by evaluating whether physical power measurements collected during embedded inference can distinguish training members from non-members without access to the model's prediction probability vector.

\section{PSCMIA Methodology}
\label{sec:methodology}

In this section, we first define the threat model considered for PSCMIA. Next, we present the overall PSCMIA workflow, including its individual profiling and attack stages.

\vspace{-0.15in}
\subsection{Threat Model}\label{subsec:threat_model}

In this paper, we consider an adversary with physical access to the target embedded device and the ability to submit a candidate input while measuring the device's power consumption during inference through external probing. The adversary knows the target model architecture, but does not require access to the target model's parameters, internal activations, gradients, training data, or prediction probability vector. The adversary has access to an auxiliary shadow dataset drawn from the same distribution as the target model's training data, but the shadow and target datasets are disjoint. The adversary uses these auxiliary datasets to train shadow models. Following the standard profiled side-channel setting, the adversary possesses a separate profiling device with the same hardware configuration as the target and can deploy shadow models on this device to collect labeled profiling traces. The adversary knows the ground-truth class label of the candidate sample, as in conventional per-class shadow-model MIA. However, it does not require the target model's predicted label.

\begin{figure*}[htp]
    \begin{center}
        \includegraphics[width=\textwidth]{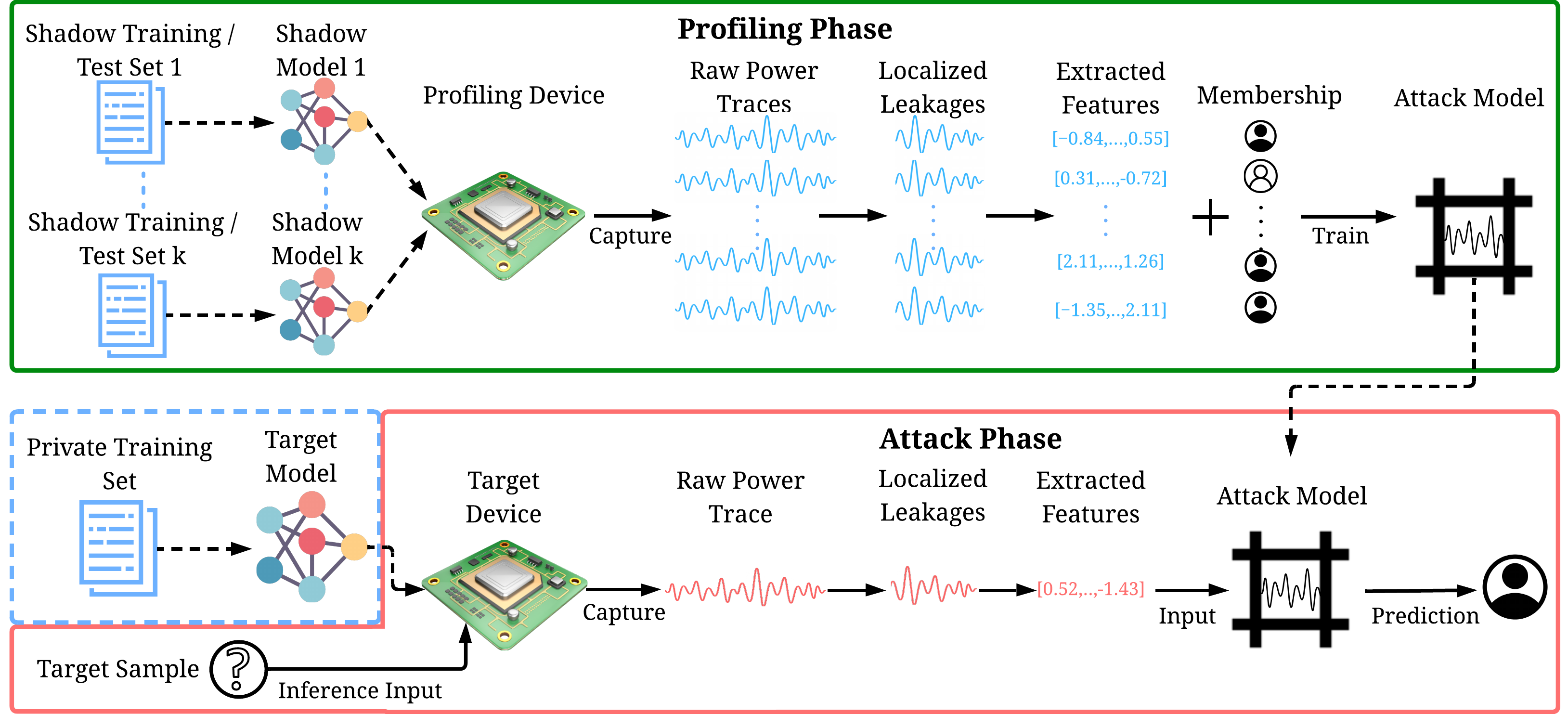}
    \end{center}
      \caption{Overview of PSCMIA that consists of two stages. During profiling, shadow models are used to construct the membership inference attack model. During the attack phase, the trained attack model is used to infer the membership of target samples.}
      \label{fig:attack_flow}
\end{figure*}

\vspace{-0.15in}
\subsection{Attack Intuition and Hypothesis}
\label{subsec:intuition}

A standard MIA is the probability vector-based shadow model attack introduced in~\cite{shokri2017membership}. In this attack, the adversary trains multiple shadow models that mimic the behavior of the target model. 
% Unlike the target model, the adversary knows the training datasets of the shadow models and, therefore, knows the membership status of each queried sample. 
The adversary queries each shadow model on samples from its own training set and on disjoint held-out samples. The resulting prediction probability vectors are labeled as \textit{in} or \textit{out} and used to train an attack model. 
% Shokri et al. formulate this process as a supervised classification problem: given a labeled record and the model's prediction vector, the attack model infers whether the record was included in the target model's training set~\cite{shokri2017membership}.

For a classification task with $C$ classes, the prediction output
of a model can be represented as a probability vector
\[
    \mathbf{p}(x) = [p_1(x), p_2(x), \ldots, p_C(x)],
\]
where $p_i(x)$ denotes the model's predicted probability that input $x$ belongs to class $i$. 
% In the original shadow-model attack, this probability vector, together with the true class label, is used as the membership signal. 
The attack exploits the observation that models often behave differently on samples used during training than on unseen samples. In particular, members are more likely to be classified correctly and often exhibit lower prediction uncertainty, higher confidence, and larger prediction margins than non-members. Prior work shows that membership inference becomes effective when the member and non-member distributions of prediction correctness and prediction uncertainty are distinguishable~\cite{shokri2017membership}.

This output-based assumption is problematic for embedded ML systems. In many embedded systems, the device exposes only the final predicted label, a control decision, or no explicit probability vector. Therefore, directly applying the original probability vector shadow model attack is not always feasible. However, the probability vector or the final prediction is not an independent artifact. It is the result of a sequence of data-dependent computations executed on the embedded processor. Thus, even when the probability vector is not externally available, the physical execution of the model may leak information correlated with the same quantities exploited by the original probability-vector shadow model MIA. We therefore formulate the following hypothesis:

\begin{tcolorbox}[highlightbox]
Power traces collected during embedded ML inference can reveal membership-dependent information correlated with the prediction behaviors used by output-based MIAs.
\end{tcolorbox}

Figure~\ref{fig:motivation_member_nonmember_property_histograms} illustrates the relationship between software-computed prediction properties and their power trace-based estimates for a  probability-vector shadow model MIA-vulnerable FC embedded ML model trained on the CINIC10 dataset~\cite{darlow2018cinic}. The software-computed properties exhibit clear differences between members and non-members, confirming that the target model exposes membership-discriminative behavior through its prediction outputs. Importantly, the corresponding power trace-based estimates preserve part of this separation, indicating that power side-channel leaks information related to these membership-relevant prediction behaviors. This observation provides the motivation for PSCMIA, which directly learns membership from inference-time power traces rather than from externally available model outputs.

We further evaluate this hypothesis in Section~\ref{subsubsec:eval_intuition}, where we provide detailed empirical evidence characterizing the membership-relevant information captured by power traces and explaining the effectiveness of PSCMIA.

\vspace{-0.15in}
\subsection{PSCMIA Overview}\label{subsec:overview}

Figure~\ref{fig:attack_flow} provides an overview of PSCMIA that consists of two main phases: a profiling phase and an attack phase. During the profiling phase, we first train the shadow models. Following the threat model defined in Section~\ref{subsec:threat_model}, we assume that the attacker has access to disjoint auxiliary datasets drawn from the same distribution as the target model's training dataset and uses these datasets to train the shadow models. A profiling device identical to the target device is then used to execute the trained shadow models and collect their power traces. We use these profiling traces for leakage localization to identify informative regions and reduce the trace length required for the attack. The selected trace segments are then processed through feature extraction, and the resulting features are used to train a binary classification attack model.

During the attack phase, the trained attack model is used to determine the membership status of a target sample from the power trace collected while the target model performs inference on the target device. The following sections describe the main activities involved in the PSCMIA workflow.

\vspace{-0.15in}
\subsection{Shadow Model Training}
\label{subsec:shadow_training}

During this stage, the adversary uses an auxiliary dataset drawn from the same distribution as the target model's training data to train multiple shadow models. The auxiliary dataset is partitioned into $k$ disjoint, equally sized subsets, with each subset containing a class-balanced set of member and non-member samples. The adversary then trains one shadow model for each partition using knowledge of the target model architecture. Each shadow model is trained on its designated member subset, while the corresponding non-member subset is held out from training. These shadow models mimic the membership-dependent behavior of the target model and provide the profiling data used in the subsequent stages.

\vspace{-0.15in}
\subsection{Leakage Localization}\label{subsec:localization}

After training each shadow model, we compile the quantized inference firmware and deploy it on the corresponding embedded device. For each input sample, the adversary sends the sample to the device and records the device's power consumption during inference. Since not all time samples in the full trace are equally informative, we use the profiling traces collected from shadow models to identify points of interest (POIs). These profiling traces have known membership labels, allowing us to locate trace regions that carry the strongest membership-relevant signal.

% The POI identification method and the resulting selected regions are presented in the next subsection.

% \subsubsection{Localization of Membership-Relevant Leakage} \label{subsubsec:poi_identification}

% The captured power trace contains some time regions that are more informative than others. Therefore, before training the final attack model, we select POIs using only the profiling traces collected from shadow models.

For each trace sample point $j$, we split the profiling traces into member and non-member groups using their known membership labels. Since the value at a single trace point is a scalar, we approximate the member and non-member distributions at that point using univariate Gaussians:
\begin{equation}
    p_{1,j} = \mathcal{N}(\mu_{1,j}, \sigma_{1,j}^{2}),
    \qquad
    p_{0,j} = \mathcal{N}(\mu_{0,j}, \sigma_{0,j}^{2}),
\end{equation}
where $p_{1,j}$ and $p_{0,j}$ denote the member and non-member trace-value distributions at sample point $j$, respectively. To ensure numerical stability, we impose a small lower bound on the estimated variances, preventing near-zero variance values from causing excessively large or undefined divergence scores.

We quantify the membership-dependent separation at that point using the symmetric KL divergence:
\begin{equation}
    D_j =
    \frac{1}{2}
    \left[
    D_{\mathrm{KL}}(p_{1,j}\Vert p_{0,j})
    +
    D_{\mathrm{KL}}(p_{0,j}\Vert p_{1,j})
    \right].
\end{equation}
We use the symmetric form because the standard KL divergence is directional, $D_{\mathrm{KL}}(p_{1,j}\Vert p_{0,j})$ and $D_{\mathrm{KL}}(p_{0,j}\Vert p_{1,j})$ can assign different scores to the same pair of distributions. Averaging the two directions gives a balanced separation score that does not depend on which distribution is treated as the reference. This is useful for POI selection because we want to identify trace points where the member and non-member distributions are  distinguishable. The score captures both mean shifts and variance differences between the two groups. 

% To localize the POIs within the inference execution, we instrument the shadow-model firmware with acquisition triggers at different layer boundaries. These triggers allow us to initiate power-trace acquisition at selected stages of model inference, such as intermediate layers and the final classifier layer. We then apply the KL-divergence-based POI selection method to the traces collected from each triggered region. This layer-wise profiling enables us to identify which internal computations expose the strongest membership-relevant side-channel signals. 

Leakage localization identifies the inference region containing strong membership-dependent separation. We then retain a fixed trace segment spanning this region rather than selecting isolated individual POIs. Section~\ref{subsubsec:poi_identification} presents the experimental evaluation of the POI identification process.

\vspace{-0.15in}
\subsection{Feature Extraction} \label{subsec:feature_extraction}

The captured power traces are high-dimensional, with each sampling point corresponding to one input dimension. To obtain a compact representation, we apply global standardization followed by principal component analysis (PCA). Specifically, a single StandardScaler is fitted jointly on all segmented traces across all classes and membership labels. For each trace sample position $j$, the profiling traces are standardized as
\begin{equation}
    \tilde{T}_{i,j} =
    \frac{T_{i,j}-\mu_j}{\sigma_j},
\end{equation}
where $\mu_j$ and $\sigma_j$ are computed exclusively from the profiling traces. We then fit PCA on the globally standardized profiling traces and retain the first 50 principal components, producing the representation $\mathbf{z}_i=\mathbf{W}_{50}^{T}\tilde{\mathbf{T}}_i$. We use 50 PCA components for all experiments without target-specific tuning. In our implementation, we use IncrementalPCA with a batch size of 128 to perform this transformation in bounded memory, since the complete high-dimensional trace matrix can be prohibitively large. Importantly, both global standardization and PCA are fitted only on the profiling partitions, and target evaluation traces are transformed using the already fitted scaler and PCA without refitting or adaptation.

The global PCA representation is shared across all classes rather than fitting a separate PCA transformation for each class. After this transformation, profiling samples are separated according to their ground-truth class to train the per-class membership classifiers. For each class, we additionally fit a class-specific StandardScaler on its 50-dimensional PCA features before training the corresponding binary attack classifier. Thus, the first scaling stage normalizes the original trace dimensions globally for PCA, whereas the second scaling stage normalizes the resulting PCA features within each class for attack-model training. Neither PCA nor either scaling stage uses the target traces during fitting, thereby preventing information leakage from the evaluation set.

\vspace{-0.15in}
\subsection{Attack Model Training and Attack}
\label{subsec:attack_model_training}

As in conventional MIAs, we formulate PSCMIA as a binary classification problem in which the attack model determines whether a candidate sample is a member of the target model’s training set. To construct the attack training set, we perform inference using the shadow models with their member and non-member samples and collect the corresponding power traces together with their membership labels. As discussed in Section~\ref{subsec:localization}, we identify POI regions that exhibit strong membership-dependent separation. However, we observed that considerable membership-related leakage is also distributed across computations preceding the highest-scoring POI region. We therefore retain a limited portion of the surrounding execution, which improves attack performance compared with using only the strongest POI region. The resulting trace segments are processed through the feature extraction stage (Section~\ref{subsec:feature_extraction}) and then used to train the attack model.

During the attack phase, a target sample whose membership status is unknown is fed into the victim inference model, and its execution power trace is collected. The collected trace is then segmented to retain the POI region identified during the profiling phase and subsequently processed through the feature-extraction step. Finally, the extracted features are provided to the trained attack model to classify the target sample as either a member or a non-member.

\section{Experimental Setup}
\label{sec:experiments}

In this section, we describe the experimental setup used to evaluate PSCMIA. We first present the target hardware platforms and power-trace acquisition setup. We then describe the target models, datasets, and attack models used in our evaluation. Finally, we introduce the baseline attacks used for comparison and define the evaluation metrics.

\vspace{-0.15in}
\subsection{Target Hardware} \label{subsec:hardware}

% We evaluate PSCMIA on two ChipWhisperer-compatible embedded platforms spanning distinct MCU architectures. Table~\ref{tab:hardware_platforms} summarizes the target hardware configurations. The platforms differ in processor architecture, clock frequency, and available Flash and SRAM, allowing us to evaluate PSCMIA under different computational and memory constraints. The STM32F3 features a 32-bit ARM Cortex-M4 core and provides a larger memory budget, supporting larger target models and larger per-class training subsets. In contrast, the XMEGA uses an 8-bit AVR architecture with limited SRAM. The XMEGA provides a non-ARM evaluation point with a different instruction set, execution characteristics, and resource budget. Evaluating PSCMIA on both platforms allows us to assess whether membership leakage through power consumption persists across heterogeneous embedded architectures.

We evaluate PSCMIA on two embedded platforms, STM32F3 and XMEGA, with distinct MCU architectures and resource constraints. The STM32F3 features a 32-bit ARM Cortex-M4 core operating at 72~MHz with 256~KB of Flash and 40~KB of SRAM, whereas the XMEGA uses an 8-bit AVR core operating at 32~MHz with 128~KB of Flash and 8~KB of SRAM. The STM32F3 provides a larger computational and memory budget, while the XMEGA offers a more resource-constrained, non-ARM evaluation point with a different instruction set and execution behavior. Evaluating both platforms allows us to assess whether PSCMIA remains effective across heterogeneous embedded architectures.

% \begin{table}[t]
% \centering

% \setlength{\tabcolsep}{3.2pt}
% \caption{Hardware platforms used to evaluate PSCMIA. }
% \label{tab:hardware_platforms}
% \small
% %\resizebox{\columnwidth}{!}{
% \begin{tabular}{llllcc}
% \toprule
% % \textbf{Platform} &
% \textbf{MCU} &
% \textbf{Core} &
% \textbf{Flash} &
% \textbf{SRAM} &
% \textbf{Clock}\\
% \midrule
% % CW Nano
% %     & STM32F0
% %     & Cortex-M0
% %     & 16 KB
% %     & 4 KB \\

% STM32F3
%     & Cortex-M4
%     & 256 KB
%     & 40 KB 
%     & 72 MHz\\

% XMEGA
%     & 8-bit AVR
%     & 128 KB
%     & 8 KB 
%     & 32 MHz\\

% \bottomrule
% \end{tabular}
% %}
% \end{table}

\vspace{-0.15in}
\subsection{Trace Acquisition}
\label{subsec:trace_collection}

We use the CW1200 ChipWhisperer-Pro and its integrated measurement infrastructure~\cite{chipwhispererpro} to acquire power traces during model inference. The ChipWhisperer-Pro was set at a sampling rate of $105$~MSamples/s and a capture buffer of 90,000 samples. During the profiling stage, we observe that the POIs are concentrated near the end of the model inference. Therefore, identifying the inference termination is sufficient to locate the relevant trace region. We construct a short power-trace template corresponding to the end-of-inference pattern using profiling traces and apply template matching to locate the inference termination in subsequent traces. The detected termination point is then used as the synchronization reference for extracting the fixed trace segment used by PSCMIA. Based on the POI localization results, we retain the final 20,000 samples for experiments on the STM32F3 and the final 10,000 samples for experiments on the XMEGA. The required trace length differs across target platforms due to differences in their computational performance and, consequently, the duration of the leakage-bearing execution region. 

To reduce measurement noise, we repeat the inference measurement five times for each input and compute the sample-wise average of the resulting traces, which is used as the final power trace. 
%Consequently, during the attack phase, PSCMIA requires five repeated queries to the target model for the target sample. 
%Thus, the query complexity of PSCMIA in our current experimental configuration is five inference executions. 
During the profiling phase, we randomize the acquisition order of member and non-member inputs before collecting their corresponding power traces. This prevents systematic acquisition-order effects from becoming correlated with membership labels.

\vspace{-0.15in}
\subsection{Target Models} \label{subsec:target_models}

We train all target and shadow models using BitNetMCU~\cite{bitnetmcu}, a low-bit quantized neural-network framework designed for resource-constrained microcontroller inference. All input images are represented as single-channel $16\times16$ images, resulting in 256 input features. We evaluate both FC and CNN architectures. The FC model follows a $256$-$128$-$64$-$32$-$10$ architecture with ReLU activations between hidden layers. The CNN uses a channel width of 64 and consists of one $3\times3$ convolution followed by two $3\times3$ depthwise convolutions, ReLU activations, and max-pooling, followed by FC layers of widths 128, 64, and 32 and a 10-class output layer.

The FC model uses 8-bit weight quantization throughout, whereas the CNN employs a mixed-precision configuration with 8-bit convolutional layers, a 2-bit symmetric first FC layer, and 8-bit quantization for the remaining FC layers. During MCU inference, low-precision activations are processed using integer arithmetic with wider intermediate accumulations and subsequent rescaling. These input-dependent arithmetic, activation, memory-access, and decision operations directly influence the measured power consumption.

\begin{table}[t]
\centering
%\footnotesize
\setlength{\tabcolsep}{2.8pt}
\caption{Overview of dataset and target models. For each dataset, we train FC and CNN BitNetMCU target models using 2,000 training samples. We report the model quantization configuration and the measured training and test accuracy.}
\label{tab:dataset_model_summary}

\resizebox{\columnwidth}{!}{%
\begin{tabular}{lc ccc ccc}
\toprule
&
&
\multicolumn{3}{c}{\textbf{FC}}
&
\multicolumn{3}{c}{\textbf{CNN}} \\
\cmidrule(lr){3-5}
\cmidrule(lr){6-8}

\textbf{Dataset}
& \textbf{Train Size}
& \textbf{Quant.}
& \textbf{Train Acc.}
& \textbf{Test Acc.}
& \textbf{Quant.}
& \textbf{Train Acc.}
& \textbf{Test Acc.} \\
\midrule

MNIST
& 2000
& 8-bit
& 100.00
& 92.15
& 8/2/8-bit
& 99.93
& 95.35 \\

FMNIST
& 2000
& 8-bit
& 99.85
& 83.23
& 8/2/8-bit
& 99.86
& 81.74 \\

CIFAR10
& 2000
& 8-bit
& 85.28
& 25.45
& 8/2/8-bit
& 83.75
& 32.35 \\

CINIC10
& 2000
& 8-bit
& 93.90
& 19.00
& 8/2/8-bit
& 78.79
& 24.13 \\

\bottomrule
\end{tabular}%
}
\vspace{-0.2in}
\end{table}

\vspace{-0.15in}
\subsection{Datasets} \label{subsec:datasets}

We evaluate the proposed attack on four image-classification datasets: MNIST~\cite{lecun1998gradient}, FashionMNIST~\cite{xiao2017fashionmnist}, CIFAR10~\cite{krizhevsky2009learning}, and CINIC10~\cite{darlow2018cinic}. MNIST contains handwritten digit images from ten classes, while FashionMNIST (FMNIST) contains ten classes of clothing and footwear images. CIFAR10 and CINIC10 are ten-class natural-image datasets, with CINIC10 providing a larger and more diverse distribution by combining CIFAR10-like classes with images drawn from ImageNet. All datasets are converted to the input format required by BitNetMCU~\cite{bitnetmcu} and used to train compact FC and CNN models.

For each dataset, we follow a shadow-model membership inference evaluation protocol. The original training data is randomly split into disjoint target training members ($D^{t}_{\mathrm{train}}$), target non-members ($D^{t}_{\mathrm{test}}$) and shadow member/non-member sets, while preserving class balance. The target model is trained on $D^{t}_{\mathrm{train}}$. The shadow models are trained using disjoint shadow training sets and evaluated using corresponding held-out shadow test sets. Since the membership status of each shadow sample is known, traces collected from the shadow models are labeled as \textit{in} or \textit{out} and used to train the attack model. In our experiments, we use four shadow models for each dataset and target configuration.

We use 200 samples per class for every dataset. For each dataset and platform, we train two target models (one FC model and one CNN model). Table~\ref{tab:dataset_model_summary} summarizes the dataset sizes, quantization, and target-model accuracies. We use fixed architecture/training configurations for the main evaluation. These configurations exhibit different levels of probability-vector membership leakage, allowing us to study the relationship between conventional and physical MIA vulnerability.

\vspace{-0.15in}
\subsection{Attack Models}\label{subsec:attack_models}

We evaluate three commonly used binary classifiers: logistic regression (LR) as a linear baseline, random forest (RF), and Extra Trees (ET) as ensemble tree classifiers. Our implementation uses 500 trees for RF and ET. The classifier configurations are kept fixed across the evaluated settings. ET was selected using profiling-validation performance and fixed before evaluation on the target models as shown in Section~\ref{subsec:pscmia_effectiveness}. 
Detailed results for all evaluated attack models are provided in Appendix~\ref{sec:apx_per_class}. 
Following the conventional per-class MIA formulation, we independently train an attack classifier for each target class and route each trace according to its input class. The classifier itself operates only on features extracted from the power trace and does not receive the target model's predicted label or output probability vector. Each shadow model contributes 2,000 member and 2,000 non-member samples, yielding $4,000 \times k$ labeled attack traces for $k$ shadow models. We use $k=4$ shadow models in our experiments, resulting in a dataset containing a total of 16,000 traces.

\vspace{-0.15in}
\subsection{Baseline Attacks}
\label{subsec:baseline_attacks}

We compare our PSCMIA against two output-based membership inference baselines: a probability vector-based shadow MIA and a label-only boundary-distance attack. These baselines represent two common black-box threat models. The shadow model attack assumes access to the model's prediction probability vector, while the label-only attack assumes access only to the predicted class label. We use these two attacks as reference points for interpreting the strength of
our PSCMIA attack. As explained in Section~\ref{subsec:intuition}, power traces partially capture output-derived properties computed from the probability vector, such as confidence, entropy, margin, and prediction correctness. Therefore, we use the shadow model attack as a higher-information reference baseline, and the label-only attack as a lower-information baseline.

\subsubsection{Shadow Model Membership Inference}
\label{subsubsec:shadow_mia}

The shadow model attack follows the standard black-box membership inference framework introduced in~\cite{shokri2017membership}. %The key idea is that a model may behave differently on samples used during training than on unseen samples. To learn this difference, the adversary trains shadow models that mimic the target model and for which the membership status of each sample is known.
We use the same four shadow models employed during the PSCMIA profiling phase, each trained on a disjoint auxiliary dataset, to ensure a fair comparison. The adversary queries each shadow model and records its prediction probability vector. These probability vectors, together with the corresponding true class labels, are used to train a binary attack classifier that predicts whether a sample is a member or non-member of the model's training set. The trained attack classifier is then evaluated against the held-out target model using its prediction probability vectors. This baseline therefore quantifies the membership information available when the target model exposes its full software-level prediction probability vector.

\subsubsection{Label-Only Membership Inference}
\label{subsubsec:label_only_mia}

The label-only baseline represents a more restrictive black-box setting in which the adversary observes only the predicted class label and does not receive prediction probabilities. 
%Prior label-only attacks exploit the observation that training samples tend to lie farther from the model's decision boundary than non-members. Therefore, membership can be inferred by estimating how much a sample must be perturbed before the model changes its predicted label.
We use a boundary-distance-based label-only attack inspired by DHAttack~\cite{li2025enhanced}. Instead of searching for the shortest adversarial perturbation, the attack measures the distance at which a sample crosses the decision boundary when it is moved toward a fixed reference point. This fixed-direction boundary distance reduces the number of queries compared with shortest-boundary-distance methods. For a target sample $x$, the adversary estimates its fixed boundary distance on the target model and compares it with the distribution of fixed boundary distances obtained from reference models trained without that sample. The resulting relative score is used as the membership signal:
\begin{equation}
    \mathrm{relScore}(x) = \mathrm{CDF}\left(d_T(x); \mathcal{G}_x\right),
\end{equation}
where $d_T(x)$ is the fixed boundary distance measured on the target model and $\mathcal{G}_x$ is the Gaussian distribution fitted to the fixed boundary distances measured on the reference models. A larger $\mathrm{relScore}(x)$ indicates that the sample behaves more like a member of the target model's training set. During our experiments, we combine the four auxiliary datasets, containing a total of 16,000 samples, to train 256 reference models following the original attack setup~\cite{li2025enhanced}. Thus, both baseline attacks use the same auxiliary-data pool as PSCMIA.

\vspace{-0.15in}
\subsection{Metrics}
\label{subsec:metrics}

We evaluate MIA as a binary classification task using ROC-AUC, false-positive rate (FPR), recall, accuracy, and TPR@1\%FPR. ROC-AUC measures the threshold-independent ability of the attack to distinguish members from non-members, where an AUC of $0.5$ corresponds to random guessing. FPR measures the fraction of non-members incorrectly classified as members, while recall (true-positive rate (TPR)) measures the fraction of true members correctly identified. Accuracy reports the overall fraction of correctly classified member and non-member samples. For RF, LR, and ET, we use the classifiers' default binary decision threshold of $0.5$ on the predicted membership probability when computing accuracy, recall, and FPR. We additionally report TPR@1\%FPR to characterize attack effectiveness at a low false-positive operating point. TPR@1\%FPR measures the fraction of members correctly identified while constraining the false-positive rate to at most $1\%$. 
% We compute this metric from the ROC curve as the maximum TPR among all operating points satisfying $\mathrm{FPR}\leq0.01$. Unlike the metrics reported at the default $0.5$ decision threshold, TPR@1\%FPR explicitly evaluates the attack under a low-FPR constraint.
When reporting a single overall metric value across all target classes, we use pooled evaluation. Specifically, we aggregate the membership labels, predicted membership labels, and attack scores of all held-out samples across classes and compute each metric once on the resulting combined set. 
% Therefore, the reported pooled metrics reflect attack performance over the complete target population rather than an average of independently computed per-class metrics.

\section{Experimental Results}
\label{sec:results}

In this section, we first present the experimental results demonstrating the effectiveness of PSCMIA. We then analyze the characteristics of the attack. Finally, we evaluate the robustness of PSCMIA through a series of ablation studies.

\vspace{-0.15in}
\subsection{Effectiveness of PSCMIA across Model Architectures and Embedded Platforms}
\label{subsec:pscmia_effectiveness}

Tables~\ref{tab:fc_mia_results} and~\ref{tab:cnn_mia_results} compare PSCMIA against probability vector-based shadow MIA and label-only MIA for FC and CNN target models deployed on the STM32F3 and XMEGA platforms. Among the three attack models explained in Section~\ref{subsec:attack_models}, ET achieves the overall best performance in profiling-validation. Therefore, we use ET as the default attack model for the remaining PSCMIA evaluations. We primarily use ROC-AUC to compare the attacks because it captures their ability to rank members above non-members independently of a particular decision threshold. Overall, PSCMIA consistently achieves ROC-AUC values above random guessing and, in several settings, approaches the performance of probability vector-based shadow MIA despite not requiring access to the target model's prediction probability vector.

\begin{table*}[t]
\centering
\scriptsize
\setlength{\tabcolsep}{3.0pt}
\caption{Membership inference attack results across datasets using FC target models. We compare probability vector-based shadow MIA, label-only MIA, and PSCMIA on STM32F3 and XMEGA. Appendix~\ref{sec:apx_per_class} provides detailed per-class results.}
\label{tab:fc_mia_results}

\resizebox{\textwidth}{!}{%
\begin{tabular}{lcccccccccccccccc}
\toprule

& \multicolumn{4}{c}{\textbf{MNIST}}
& \multicolumn{4}{c}{\textbf{FMNIST}}
& \multicolumn{4}{c}{\textbf{CIFAR10}}
& \multicolumn{4}{c}{\textbf{CINIC10}} \\

\cmidrule(lr){2-5}
\cmidrule(lr){6-9}
\cmidrule(lr){10-13}
\cmidrule(lr){14-17}

\textbf{Metric}
& \multirow{2}{*}{\textbf{\shortstack[c]{Shadow\\Model}}}
& \multirow{2}{*}{\textbf{\shortstack[c]{Label\\Only}}}
& \multicolumn{2}{c}{\textbf{PSCMIA}}
& \multirow{2}{*}{\textbf{\shortstack[c]{Shadow\\Model}}}
& \multirow{2}{*}{\textbf{\shortstack[c]{Label\\Only}}}
& \multicolumn{2}{c}{\textbf{PSCMIA}}
& \multirow{2}{*}{\textbf{\shortstack[c]{Shadow\\Model}}}
& \multirow{2}{*}{\textbf{\shortstack[c]{Label\\Only}}}
& \multicolumn{2}{c}{\textbf{PSCMIA}}
& \multirow{2}{*}{\textbf{\shortstack[c]{Shadow\\Model}}}
& \multirow{2}{*}{\textbf{\shortstack[c]{Label\\Only}}}
& \multicolumn{2}{c}{\textbf{PSCMIA}} \\

\cmidrule(lr){4-5}
\cmidrule(lr){8-9}
\cmidrule(lr){12-13}
\cmidrule(lr){16-17}

&
&
& \textbf{STM32F3} & \textbf{XMEGA}
&
&
& \textbf{STM32F3} & \textbf{XMEGA}
&
&
& \textbf{STM32F3} & \textbf{XMEGA}
&
&
& \textbf{STM32F3} & \textbf{XMEGA} \\

\midrule

ROC-AUC
& 0.6582 & 0.5281 & 0.6193 & 0.6186
& 0.6804 & 0.6016 & 0.6795 & 0.6477
& 0.8794 & 0.8137 & 0.7958 & 0.7952
& 0.9445 & 0.8960 & 0.9065 & 0.8925\\

FPR
& 0.7090 & 0.3185 & 0.1010 & 0.4885
& 0.7215 & 0.4715 & 0.4340 & 0.5330
& 0.1920 & 0.2355 & 0.2540 & 0.2435
& 0.1365 & 0.2000 & 0.1445 & 0.1760\\

Recall
& 0.9365 & 0.3545 & 0.1505 & 0.6505
& 0.9775 & 0.6260 & 0.6815 & 0.7445
& 0.8010 & 0.7795 & 0.6870 & 0.7215
& 0.8930 & 0.9035 & 0.8060 & 0.8415\\

TPR@1\%FPR
& 0.0155 & 0.0305 & 0.0140 & 0.0300
& 0.0265 & 0.0515 & 0.0365 & 0.0145
& 0.1635 & 0.0905 & 0.0675 & 0.0885
& 0.3925 & 0.1730 & 0.1965 & 0.2180\\

Acc.
& 0.6138 & 0.5180 & 0.5248 & 0.5810
& 0.6280 & 0.5773 & 0.6238 & 0.6058
& 0.8045 & 0.7720 & 0.7165 & 0.7390
& 0.8782 & 0.8518 & 0.8308 & 0.8328\\

\bottomrule
\end{tabular}%
}
\end{table*}

\begin{table*}[t]
\centering
\scriptsize
\setlength{\tabcolsep}{3.0pt}
\caption{Membership inference attack results across datasets using CNN target models. We compare probability vector-based shadow MIA, label-only MIA, and PSCMIA on STM32F3 and XMEGA. Appendix~\ref{sec:apx_per_class} provides detailed per-class results.}
\label{tab:cnn_mia_results}

\resizebox{\textwidth}{!}{%
\begin{tabular}{lcccccccccccccccc}
\toprule

& \multicolumn{4}{c}{\textbf{MNIST}}
& \multicolumn{4}{c}{\textbf{FMNIST}}
& \multicolumn{4}{c}{\textbf{CIFAR10}}
& \multicolumn{4}{c}{\textbf{CINIC10}} \\

\cmidrule(lr){2-5}
\cmidrule(lr){6-9}
\cmidrule(lr){10-13}
\cmidrule(lr){14-17}

\textbf{Metric}
& \multirow{2}{*}{\textbf{\shortstack[c]{Shadow\\Model}}}
& \multirow{2}{*}{\textbf{\shortstack[c]{Label\\Only}}}
& \multicolumn{2}{c}{\textbf{PSCMIA}}
& \multirow{2}{*}{\textbf{\shortstack[c]{Shadow\\Model}}}
& \multirow{2}{*}{\textbf{\shortstack[c]{Label\\Only}}}
& \multicolumn{2}{c}{\textbf{PSCMIA}}
& \multirow{2}{*}{\textbf{\shortstack[c]{Shadow\\Model}}}
& \multirow{2}{*}{\textbf{\shortstack[c]{Label\\Only}}}
& \multicolumn{2}{c}{\textbf{PSCMIA}}
& \multirow{2}{*}{\textbf{\shortstack[c]{Shadow\\Model}}}
& \multirow{2}{*}{\textbf{\shortstack[c]{Label\\Only}}}
& \multicolumn{2}{c}{\textbf{PSCMIA}} \\

\cmidrule(lr){4-5}
\cmidrule(lr){8-9}
\cmidrule(lr){12-13}
\cmidrule(lr){16-17}

&
&
& \textbf{STM32F3} & \textbf{XMEGA}
&
&
& \textbf{STM32F3} & \textbf{XMEGA}
&
&
& \textbf{STM32F3} & \textbf{XMEGA}
&
&
& \textbf{STM32F3} & \textbf{XMEGA} \\

\midrule

ROC-AUC
& 0.6089 & 0.5111 & 0.5821 & 0.5784
& 0.6945 & 0.5681 & 0.6514 & 0.6588
& 0.7726 & 0.7285 & 0.7464 & 0.7668
& 0.8572 & 0.7978 & 0.7417 & 0.7757 \\

FPR
& 0.5040 & 0.1040 & 0.0560 & 0.5165
& 0.6880 & 0.2300 & 0.1370 & 0.4875
& 0.3170 & 0.2420 & 0.2360 & 0.3010
& 0.2290 & 0.2125 & 0.2350 & 0.2995 \\

Recall
& 0.6420 & 0.1405 & 0.0680 & 0.6060
& 0.9535 & 0.3750 & 0.2460 & 0.7010
& 0.7190 & 0.6210 & 0.5350 & 0.7490
& 0.7800 & 0.7265 & 0.5770 & 0.7315 \\

TPR@1\%FPR
& 0.0120 & 0.0235 & 0.0140 & 0.0240
& 0.0300 & 0.0455 & 0.0360 & 0.0215
& 0.0780 & 0.0870 & 0.0540 & 0.0050
& 0.0880 & 0.1315 & 0.0405 & 0.0475 \\

Acc.
& 0.5690 & 0.5182 & 0.5060 & 0.5448
& 0.6328 & 0.5725 & 0.5545 & 0.6068
& 0.7010 & 0.6895 & 0.6495 & 0.7240
& 0.7755 & 0.7570 & 0.6710 & 0.7160 \\

\bottomrule
\end{tabular}%
}
\end{table*}

\begin{table*}[t]
\centering
\small
\setlength{\tabcolsep}{3.2pt}
\caption{Discrete probability vector-property recovery from power traces. The first row reports a probability vector-based shadow model MIA sanity check. The remaining rows report performance for classifiers trained to recover discrete probability vector-derived properties only from power traces. Here, \textit{Base Acc.} denotes the accuracy of always predicting the most frequent held-out class. Base BAcc. is $1/K$, where $K$ is the number of target classes observed in both profiling and held-out evaluation.}
\label{tab:discrete_output_property_recovery}
\begin{tabular}{lccccccccc}
\toprule
\textbf{Target property} &
\textbf{Model} &
\textbf{Classification} &
\textbf{Base Acc.} &
\textbf{Acc.} &
\textbf{Base BAcc.} &
\textbf{BAcc.} &
\textbf{F1} &
\textbf{ROC-AUC} &
\textbf{Top-2 Acc.} \\
\midrule
Probability-vector Shadow Model MIA
    & LR
    & Membership
    & 0.500 & 0.878
    & 0.500 & 0.878
    & 0.880 & 0.944
    & -- \\

\midrule
Class prediction
    & LR
    & 10 classes
    & 0.111 & 0.522
    & 0.100 & 0.520
    & 0.517 & 0.929
    & 0.805 \\

Class prediction
    & ET
    & 10 classes
    & 0.111 & \textbf{0.622}
    & 0.100 & \textbf{0.620}
    & 0.617 & \textbf{0.955}
    & 0.872 \\

\midrule
Correctness prediction
    & LR
    & Binary
    & 0.565 & 0.593
    & 0.500 & 0.594
    & 0.591 & 0.637
    & -- \\

Correctness prediction
    & ET
    & Binary
    & 0.565 & \textbf{0.616}
    & 0.500 & \textbf{0.598}
    & 0.596 & \textbf{0.650}
    & -- \\

\midrule
Top-1 probability bin
    & LR
    & 8 classes
    & 0.194 & 0.193
    & 0.125 & \textbf{0.283}
    & 0.180 & \textbf{0.698}
    & 0.428 \\

Top-1 probability bin
    & ET
    & 8 classes
    & 0.194 & \textbf{0.275}
    & 0.125 & 0.249
    & 0.264 & 0.693
    & 0.511 \\

\midrule
Margin bin
    & LR
    & 9 classes
    & 0.197 & 0.187
    & 0.111 &\textbf{ 0.209}
    & 0.160 & \textbf{0.661}
    & 0.363 \\

Margin bin
    & ET
    & 9 classes
    & 0.197 & \textbf{0.220}
    & 0.111 & 0.180
    & 0.177 & 0.655
    & 0.405 \\

\bottomrule
\end{tabular}
\end{table*}

For FC models on STM32F3, PSCMIA achieves ROC-AUC values of $0.6193$, $0.6795$, $0.7958$, and $0.9065$ on MNIST, FMNIST, CIFAR10, and CINIC10, respectively. The best result is observed on CINIC10, where PSCMIA reaches an ROC-AUC of $0.9065$, compared with $0.9445$ for shadow MIA. On FMNIST, the difference is even smaller, with PSCMIA achieving $0.6795$ compared with $0.6804$ for shadow MIA. These results show that the power side-channel preserves the membership-related information available through the model's output probability vector.

PSCMIA also compares favorably with the more restrictive label-only baseline. For FC models, PSCMIA improves ROC-AUC over label-only MIA by $0.0912$ on MNIST and $0.0779$ on FMNIST, and slightly improves the ROC-AUC on CINIC10 from $0.8960$ to $0.9065$. CIFAR10 is the only FC setting in which label-only MIA performs better, achieving an ROC-AUC of $0.8137$ compared with $0.7958$ for PSCMIA. Importantly, these results are achieved using only power traces collected during inference, without access to either the prediction probability vector or even the predicted label. In contrast, the label-only baseline explicitly assumes that the adversary can observe the target model's prediction label. Therefore, PSCMIA demonstrates that physical execution can expose membership information even when the model interface reveals no prediction output to the adversary. 
% \textcolor{blue}{The variation in attack performance across datasets indicates that the extent of membership information observable through the power side channel depends on the target model and dataset.}

A similar trend is observed for CNN models. PSCMIA achieves ROC-AUC values of $0.5821$, $0.6514$, $0.7464$, and $0.7417$ on MNIST, FMNIST, CIFAR10, and CINIC10, respectively. Although shadow MIA remains the strongest baseline, with corresponding ROC-AUC values of $0.6089$, $0.6945$, $0.7726$, and $0.8572$, the gap between PSCMIA and shadow MIA remains relatively small: $0.0268$, $0.0431$, $0.0262$, and $0.1155$, respectively. More importantly, PSCMIA outperforms the label-only baseline on MNIST, FMNIST, and CIFAR10 CNN models, while label-only MIA performs slightly better on CINIC10. The ROC-AUC improvement is $0.0710$ for MNIST, $0.0833$ for FMNIST, and $0.0179$ for CIFAR10. This consistency across both FC and CNN architectures indicates that the observed membership leakage is not specific to a single model architecture.

Attack results on XMEGA for both FC and CNN target models exhibit trends consistent with those observed on STM32F3. This consistency across two distinct MCU architectures provides evidence that PSCMIA is not specific to a single hardware platform and remains effective across different target models, datasets, and embedded devices.

% For both model architectures, PSCMIA achieves its strongest performances on the CINIC10 and CIFAR10 datasets. Notably, CINIC10 is the dataset for which the target models exhibit the highest vulnerability to probability vector-based shadow MIA. This observation is consistent with the intuition presented in Section~\ref{subsec:intuition}, where power side-channel traces capture information related to prediction behaviors that are also exploited by output-based MIAs like shadow MIA. 
These results show that target models exhibiting stronger membership-related behavior through their prediction probability vectors also tend to exhibit stronger membership leakage through power side-channel. Moreover, the variation across datasets indicates that power-based membership leakage is influenced by the characteristics of the trained target model rather than being uniformly present across all inference executions. We further investigate the factors contributing to this variation in the subsequent analysis.

\vspace{-0.15in}
\subsection{Characterization of PSCMIA}

This section first provides empirical evidence supporting the hypothesis introduced in Section~\ref{subsec:intuition}. We then experimentally evaluate the POI identification process used to localize the membership-relevant leakage regions in the power traces.

\subsubsection{Analysis of Power Side-Channel-Based Membership Leakage}
\label{subsubsec:eval_intuition}

To examine the hypothesis introduced in Section~\ref{subsec:intuition}, we investigate whether inference-time power traces contain information correlated with prediction correctness and uncertainty. We use the FC model trained on CINIC10 and deployed on STM32F3 for this analysis.

% , as this configuration exhibits the highest baseline membership leakage among the evaluated target models. 
% The dataset is partitioned into five disjoint subsets. Each subset contains 2,000 member images and 2,000 non-member images, balanc ed across the ten classes with 200 images per class. We train one model per subset, use one model as the target model, and use the remaining four models as shadow models. The selected target model achieves 93.9\% training accuracy and 19.0\% test accuracy. A detailed discussion on datasets used models can be found in Section~\ref{sec:experiments}

We first record the software probability vectors produced by the corresponding Shadow models and use them to perform the probability vector-based shadow model MIA. This step serves as a sanity check, verifying that the selected target model exhibits measurable membership leakage through its output probabilities. We use the CINIC10-FC configuration as the case study for this characterization because it exhibits the strongest probability-vector Shadow model MIA performance among the evaluated configurations.

After establishing this baseline vulnerability, we collect power traces during inference on the STM32F3 for both member and non-member inputs of each shadow model. Then the previously recorded probability vectors are used to compute prediction-behavior quantities that are commonly exploited by output-based MIAs, including prediction correctness, confidence, normalized entropy, and prediction margin. Prediction correctness indicates whether the predicted label matches the ground-truth label. Confidence is the maximum prediction probability and normalized entropy measures the uncertainty of the prediction probability vector. Prediction margin is the difference between the largest and second-largest prediction probabilities.
% \begin{equation}
%     \mathrm{correctness}(x)
%     =
%     \mathbb{I}\!\left[\arg\max_{c} p_c(x) = y\right],
% \end{equation}
% \begin{equation}
%     \mathrm{confidence}(x) = \max_{c \in \{1,\ldots,C\}} p_c(x),
% \end{equation}
% \begin{equation}
%     \mathrm{entropy_{norm}}(x)
%     =
%     -\frac{1}{\log C}
%     \sum_{c=1}^{C} p_c(x)\log p_c(x),
% \end{equation}
% \begin{equation}
%     \mathrm{margin}(x) = p_{(1)}(x) - p_{(2)}(x),
% \end{equation}
% where $x$ denotes the model input, $y$ is its ground-truth class label, and $C$ is the number of classes. The quantity $p_c(x)$ denotes the predicted probability assigned to class $c$. Furthermore, $p_{(1)}(x)$ and $p_{(2)}(x)$ denote the largest and second-largest predicted probabilities, respectively.
% The correctness metric is binary, taking a value of $1$ when the predicted class matches the ground-truth label and $0$ otherwise.

Since the target model is already vulnerable to shadow-model MIA when probability vectors are available, we next ask whether the same membership-relevant properties can be inferred from power traces. Specifically, we train classifiers to predict correctness, confidence level, entropy level, and margin level using only side-channel measurements. If these quantities can be predicted from power traces, it provides evidence that physical power side-channel measurements capture information correlated with the same prediction-behavior signals used by output-based membership inference attacks. This analysis provides an empirical explanation for the PSCMIA effectiveness observed in Section~\ref{subsec:pscmia_effectiveness}.

% \begin{table}[t]
% \centering
% \small
% \setlength{\tabcolsep}{5.2pt}
% \caption{Regression performance for probes trained to estimate probability vector-derived continuous properties from power traces. Lower MAE and RMSE indicate better recovery, while higher $R^2$ and Spearman correlation $\rho$ indicate stronger agreement with the software-computed property.}
% \label{tab:continuous_output_property_recovery}
% \begin{tabular}{llcccc}
% \toprule
% \textbf{Property} & \textbf{Probe} & \textbf{MAE} & \textbf{RMSE} & \textbf{$R^2$} & \textbf{$\rho$} \\
% \midrule
% Confidence             & Ridge       & 0.125 & 0.150 & 0.258 & 0.497 \\
%                        & Ex. Trees & 0.119 & 0.144 & 0.318 & 0.560 \\
% \midrule
% Entropy                & Ridge       & 0.237 & 0.297 & 0.224 & 0.464 \\
%                        & Ex. Trees & 0.223 & 0.281 & 0.304 & 0.553 \\
% \midrule
% Prob. margin     & Ridge       & 0.176 & 0.208 & 0.230 & 0.461 \\
%                        & Ex. Trees & 0.168 & 0.202 & 0.279 & 0.513 \\
% \midrule
% Logit margin           & Ridge       & 0.600 & 0.758 & 0.244 & 0.440 \\
%                        & Ex. Trees & 0.577 & 0.736 & 0.286 & 0.489 \\
% \midrule
% Class prob.            & Ridge       & 0.235 & 0.316 & 0.126 & 0.344 \\
%                        & Ex. Trees & 0.227 & 0.310 & 0.158 & 0.391 \\
% \bottomrule
% \end{tabular}
% \end{table}

Table~\ref{tab:discrete_output_property_recovery} and Figure~\ref{fig:motivation_member_nonmember_property_histograms} provide a chain of evidence for our motivating hypothesis. As shown in the first row of Table~\ref{tab:discrete_output_property_recovery}, the probability vector-based Shadow model MIA achieves ROC-AUC of $0.944$ and balanced accuracy of $0.878$, confirming that the software probability vector contains strong membership information. This serves as a sanity check before evaluating whether similar information is exposed through physical side-channel measurements.

After the MIA sanity check, the remaining rows of Table~\ref{tab:discrete_output_property_recovery} evaluate whether power traces contain information on discrete output-derived properties that are commonly associated with membership leakage. Predicted-class recovery reaches $0.622$ accuracy and $0.620$ balanced accuracy using ET, compared to a $0.111$ majority-class baseline and $0.100$ balanced-accuracy baseline. This indicates that the trace contains substantial information about the model's decision behavior. Correctness prediction also exceeds the constant baseline, achieving $0.650$ ROC-AUC with ET. Since prediction correctness is one of the factors that separates members from non-members, this result suggests that traces capture part of the membership-relevant execution behavior.

The bin-wise recovery tasks provide a finer-grained view of confidence- and margin-related leakage. For top-1 0.1-probability bins, LR improves balanced accuracy from the $0.125$ baseline (8 bins from 0.1-0.2 to 0.8-0.9) to $0.283$ and achieves $0.698$ macro-averaged one-vs-rest ROC-AUC. For 0.1-margin bins, logistic regression improves balanced accuracy from
$0.111$ (9 bins from 0.0-0.1 to 0.8-0.9) to $0.209$ and achieves $0.661$ ROC-AUC. Although the raw accuracy for some bin-wise tasks is affected by class imbalance, the balanced-accuracy and AUC improvements show that the probes learn non-trivial information about the output-probability distribution from the power trace. 
A detailed comparison of PSCMIA with several existing output-based MIA approaches is provided in Appendix~\ref{sec:apx_per_class}, in Tables~\ref{tab:fc_mia_results_full_sota} and~\ref{tab:cnn_mia_results_full_sota}.

\begin{figure*}[t]
    \centering

    % ==========================================================
    % Row 1: Shared legend
    % ==========================================================
    \includegraphics[
        width=0.7\textwidth
    ]{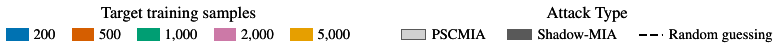}

    % ==========================================================
    % Row 2: Two figures side by side
    % ==========================================================
    \begin{subfigure}[t]{0.49\textwidth}
        \centering
        \includegraphics[width=\linewidth]{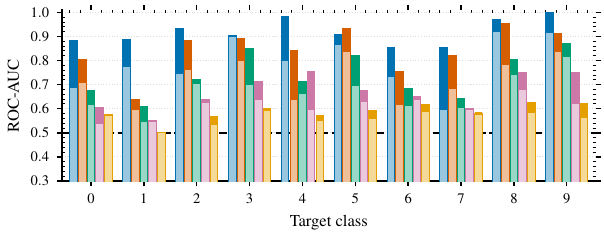}
        \caption{MNIST-FC XMEGA.}
        \label{fig:mnist_pro_training_size}
    \end{subfigure}
    \hfill
    \begin{subfigure}[t]{0.49\textwidth}
        \centering
        \includegraphics[width=\linewidth]{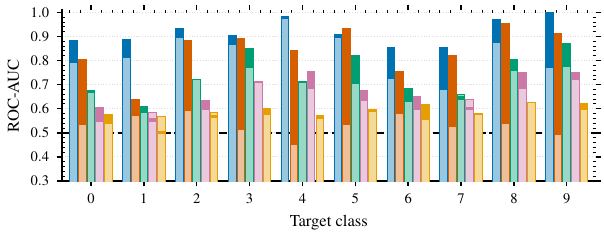}
        \caption{MNIST-FC STM32F3.}
        \label{fig:mnist_nano_training_size}
    \end{subfigure}

    \caption{
    PSCMIA attack ROC-AUC across increasing target-model training dataset sizes, compared with Shadow-MIA.}
    \label{fig:training_size_sweep_mnist}

\end{figure*}

\subsubsection{Localization of Membership-Relevant Leakage} \label{subsubsec:poi_identification}

\begin{figure}[t]
\centering
\begin{subfigure}{\columnwidth}
    \centering
    \includegraphics[width=\textwidth]{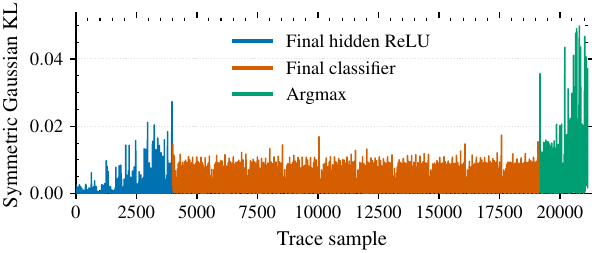}
    \caption{CINIC10-FC STM32F3.}
    \label{fig:cinic10_pro_poi}
\end{subfigure}

\begin{subfigure}{\columnwidth}
    \centering
     \includegraphics[width=\textwidth]{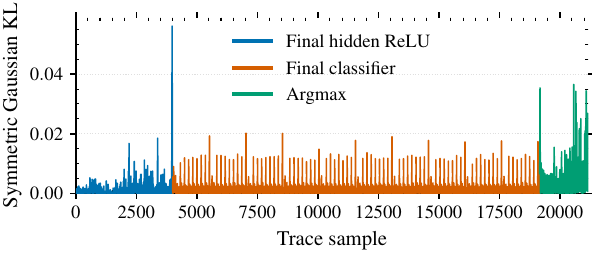}
    \caption{CIFAR10-CNN STM32F3}
    \label{fig:cifar10_pro_poi}
\end{subfigure}
\caption{Layer-wise localization of membership-relevant POI using symmetric KL divergence. This shows that most of the POIs are in the region related to the argmax operation.}
\label{fig:layerwise_poi_results}
\vspace{-0.15in}
\end{figure}

Figure~\ref{fig:layerwise_poi_results} presents the layer-wise POI localization results for both FC and CNN models. Figure~\ref{fig:cinic10_pro_poi} depicts the FC model trained on CINIC10 and executed on the STM32F3 target. Larger KL-divergence values indicate stronger separation between member and non-member profiling traces. This result shows that the most informative POIs are concentrated near the final argmax stage. This stage processes the final class scores to determine the predicted label and therefore reflects output decision behavior that is closely associated with the membership-related properties discussed in Section~\ref{subsec:intuition}. 
% In comparison, the preceding ReLU and final fully connected regions exhibit lesser member/non-member separation.
A similar localization pattern is observed for the CIFAR10 dataset with CNN architecture. Although the exact POI locations vary across datasets and model architectures, the strongest membership-relevant leakage generally appears near the final stages of inference. These results indicate that the final output computation and decision logic expose more discriminative membership-related side-channel information than earlier stages. 

To determine the power-trace region required for PSCMIA, we evaluate the attack using combinations of the final inference operations. We perform PSCMIA on CINIC10-FC STM32F3 using only the identified POI region corresponding to the argmax operation and it yields an ROC-AUC of $0.8732$. Extending the trace to include the final classifier and {argmax} results in an ROC-AUC of $0.8051$, while including the final hidden ReLU layer, final classifier, and {argmax} achieves the highest ROC-AUC of $0.9065$. These results indicate that the membership-related leakage is concentrated near the end of inference, but is distributed across multiple terminal operations rather than confined to a single instruction region.

Based on this observation, we use a fixed terminal segment immediately preceding the completion of inference for all subsequent experiments. We retain the final 20,000 samples for STM32F3 and the final 10,000 samples for XMEGA, with the platform-specific lengths selected to cover the identified leakage-bearing operations. This reduces the trace requirement from the entire inference execution to a compact trace segment while maintaining strong attack performance. 

\vspace{-0.15in}
\subsection{Ablation and Robustness Studies}

\subsubsection{Impact of Target-Model Overfitting on PSCMIA}

In this section, we investigate how target-model overfitting influences PSCMIA effectiveness. As shown in Figure~\ref{fig:training_size_sweep_mnist}, PSCMIA generally achieves higher per-class ROC-AUC when the target model is trained with smaller training sets, where the corresponding train-test accuracy gap is larger. Table~\ref{tab:overfitting_mia} presents the target models' training and test accuracies together with the corresponding training-set sizes. As the training set size increases and the target model generalizes better, the attack ROC-AUC decreases toward random-guessing performance for most classes. This trend is consistently observed across both embedded platforms, indicating that the relationship is not specific to a particular MCU architecture. However, a similar trend is also observed for the probability vector-based Shadow MIA, further supporting the relationship between PSCMIA effectiveness and the target model's underlying vulnerability to conventional MIA. These results show that target-model overfitting is an important factor in PSCMIA effectiveness.

\begin{table}[t]
\centering
\scriptsize
\setlength{\tabcolsep}{3.0pt}
\caption{Effect of target-model training-set size and generalization gap on MIA effectiveness for the MNIST FC model.}
\label{tab:overfitting_mia}
\begin{tabular}{lcccccc}
\toprule
\textbf{Train} &
\textbf{Train} &
\textbf{Test} &
\textbf{Gen.} &
\textbf{Shadow MIA} &
\multicolumn{2}{c}{\textbf{PSCMIA Pooled ROC-AUC}} \\
\cmidrule(lr){6-7}
\textbf{Size} &
\textbf{Acc. (\%)} &
\textbf{Acc. (\%)} &
\textbf{Gap} &
\textbf{Pooled AUC} &
\textbf{STM32F3} &
\textbf{XMEGA} \\
\midrule
200   & 100.00 & 70.40 & 29.60 & 0.8974 & 0.7936 & 0.7918 \\
500   &  99.96 & 83.68 & 16.28 & 0.8345 & 0.7095 & 0.7442 \\
1000  &  99.90 & 89.70 & 10.20 & 0.7238 & 0.6699 & 0.6768 \\
2000  & 100.00 & 92.15 &  7.85 & 0.6582 & 0.6193 & 0.6186 \\
5000  & 100.00 & 94.94 &  5.06 & 0.5681 & 0.5680 & 0.5632 \\
\bottomrule
\end{tabular}
\end{table}

% \color{blue}
% The results are consistent with the underlying intuition of PSCMIA where an overfit model exhibits more distinguishable behavior between samples encountered during training and previously unseen samples, which in turn produces stronger membership-dependent information in its power side-channel. The magnitude of this leakage varies across target classes, and the decrease in ROC-AUC is not strictly monotonic for every class. These results show that target-model overfitting is an important factor in PSCMIA effectiveness.
% \color{black}

\subsubsection{Impact of Model Prediction Correctness}

As shown in Table~\ref{tab:fc_mia_results}, the CINIC10-FC model exhibits the highest vulnerability to PSCMIA. Together with the observations in Section~\ref{subsubsec:eval_intuition} and CINIC10-FC model's testing accuracy, this suggests that differences in prediction correctness between members and non-members may contribute to the attack's effectiveness. To isolate this effect, we performed an additional experiment using only correctly classified samples. During profiling, we retain power traces corresponding to correctly classified member and non-member samples and create a new balanced dataset. Then we train the attack model exclusively on this subset. The same restriction is applied during evaluation.

When all samples are considered, PSCMIA achieves an ROC-AUC of $0.9065$ on the CINIC10-FC model deployed on STM32F3. Restricting the attack to correctly classified samples reduces the ROC-AUC to $0.6851$, with an accuracy of $0.6295$ and an FPR of $0.0308$. Similarly, for CIFAR10-FC on STM32F3, ROC-AUC decreases from $0.7958$ to $0.6030$. This substantial reduction confirms that prediction correctness contributes significantly to the membership-related information captured by PSCMIA. However, the ROC-AUC remains above random guessing even after correctness is held constant. This indicates that prediction correctness alone does not explain the observed leakage; PSCMIA also captures additional membership-correlated characteristics of the model's inference-time execution.

\subsubsection{Attack Model Transferability Across Datasets}

\begin{figure}[htp]
    \begin{center}
        \includegraphics[width=\columnwidth]{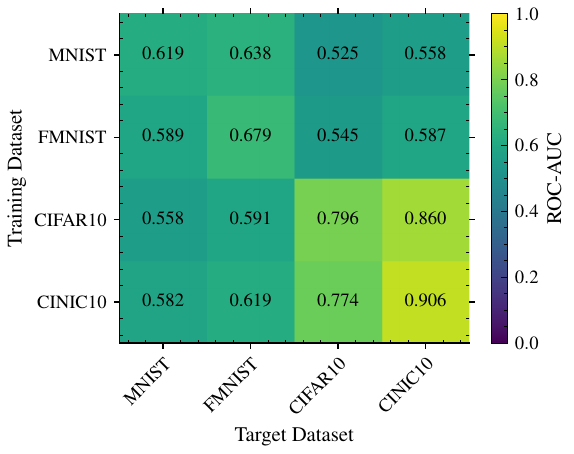}
    \end{center}
      \caption{Transferability of PSCMIA across datasets on STM32F3. Rows indicate the source dataset used to train, while columns indicate the target dataset used for evaluation.}
      \label{fig:attack_transferability}
      \vspace{-0.15in}
\end{figure}

% \begin{figure*}[hpt]
% \centering
% \begin{subfigure}{\columnwidth}
% \centering
% \includegraphics[width=\textwidth]{fig/logreg_roc_auc_transfer_heatmap.pdf}
% \vspace{-0.2in}
% \caption{ROC-AUC of attack under cross-dataset and cross-device transfer.}
% \label{fig:transferability_auc}
% \end{subfigure}
% \hfill
% \begin{subfigure}{\columnwidth}
% \centering
% \includegraphics[width=\textwidth]{fig/logreg_accuracy_transfer_heatmap.pdf}
% \vspace{-0.2in}
% \caption{Accuracy of attack under cross-dataset and cross-device transfer.}
% \label{fig:transferability_accuracy}
% \end{subfigure}
% %\vspace{-0.2in}
% \caption{Transferability of PSCMIA across datasets. Rows indicate the source dataset used to train the class-specific attack models, while columns indicate the target dataset used for evaluation. Diagonal entries correspond to within-dataset evaluation, whereas off-diagonal entries quantify cross-dataset transferability.}
% \label{fig:attack_transferability}
% %\vspace{-0.15in}
% \end{figure*}

Figure~\ref{fig:attack_transferability} shows that PSCMIA attack models achieve the strongest performance when the profiling and target datasets match, but meaningful cross-dataset transfer is still possible. The most notable result is observed between CIFAR10 and CINIC10. An attack model trained on CIFAR10 achieves an ROC-AUC of $0.860$ when evaluated on CINIC10, compared with $0.906$ ROC-AUC for within-CINIC10 evaluation. Transfer in the reverse direction also remains effective, with CINIC10 to CIFAR10 achieving an ROC-AUC of $0.774$ compared with $0.796$ for within-CIFAR10 evaluation. Notably, CIFAR10 and CINIC10 are also the two datasets that exhibit the highest vulnerability across all three evaluated MIA approaches. This higher underlying membership leakage may partly explain the strong transferability observed between them. In addition, CIFAR10 and CINIC10 contain visually similar image classes and related input characteristics, which likely induce more similar inference behavior and power-leakage patterns. MNIST to FMNIST transfer achieves an ROC-AUC of $0.638$, close to the within FMNIST result of $0.679$. These results indicate that PSCMIA can preserve a substantial portion of its membership-related capability when the source and target datasets exhibit similar characteristics. 
Appendix~\ref{sec:apx_transferability} provides additional transferability results for CNN models and the XMEGA platform.

In contrast, transfer between more dissimilar datasets generally results in ROC-AUC values closer to random guessing, indicating that the learned attack representation is not completely dataset independent. Overall, the results suggest that PSCMIA captures both dataset-specific and transferable membership-related characteristics from power traces. However, the degradation observed across dissimilar datasets shows that similarity between the profiling and target settings remains an important factor in attack effectiveness.

\section{Applicability and Limitations}
\label{sec:limits}

While PSCMIA is applicable across datasets (MNIST, FMNIST, CIFAR10, CINIC10), ML models (FC and CNN), and embedded platforms (STM32F3, XMEGA), it has several limitations. First, the current attack operates on a per-class basis and assumes that the adversary knows the ground-truth class label of the target sample. Removing this assumption and developing class-agnostic attack models remains an important direction for future work. Second, PSCMIA relies on physically measured power traces and therefore requires physical access to the target device during the attack phase, as well as access to a sufficiently similar profiling device for constructing the attack model. Restricting physical access or preventing representative profiling can consequently reduce the practicality of the attack.

Our experiments also show that PSCMIA effectiveness is correlated with the target model's underlying susceptibility to membership inference, with models that are more vulnerable to probability vector-based shadow MIA generally exhibiting stronger power-based membership leakage. Therefore, PSCMIA should not be expected to achieve uniformly high effectiveness against models with inherently weak membership leakage. However, Appendix~\ref{sec:apx_defense} shows that PSCMIA retains considerable effectiveness even after removing the trace segment corresponding to the argmax computation. This indicates that defenses that suppress or remove leakage from the identified POI region are unlikely to fully eliminate PSCMIA, since membership-relevant information can remain distributed across other stages of inference. Finally, our evaluation focuses on image-classification workloads, where extending PSCMIA to non-image modalities and other ML tasks remains an important direction for future work.

\section{Conclusion}\label{sec:conclusion}

In this paper, we presented PSCMIA: a power side-channel-based membership inference attack against embedded ML models. Unlike conventional MIAs that rely on prediction probability vectors or predicted labels, PSCMIA infers membership solely from physically measured power traces collected during model inference. Our key observation is that restricting access to model outputs does not necessarily eliminate membership leakage, since the data-dependent computations that produce these outputs are physically executed by the underlying processor and can be reflected in its power consumption. We demonstrated this leakage across multiple datasets, different ML models, and embedded MCU platforms. PSCMIA achieves ROC-AUC values of up to $0.9065$. Moreover, PSCMIA outperforms label-only MIA in most evaluated configurations while approaching the performance of the shadow MIA in several configurations.

Our analysis further provides insight into when and why PSCMIA is effective. Target models that exhibit stronger vulnerability to probability vector-based shadow MIA generally show greater susceptibility to PSCMIA, indicating a relationship between software-visible membership behavior and physical membership leakage. We further show that target-model overfitting and prediction correctness influence attack effectiveness, and that PSCMIA can transfer across related datasets. These results demonstrate that physically accessible embedded ML systems introduce a membership-privacy attack surface that is not captured by  output-based threat models. Therefore, protecting sensitive on-device ML deployments requires considering not only what information is exposed through the software interface, but also what information is unintentionally revealed through their physical side-channels.

\ifCLASSOPTIONcompsoc
  \section*{Code Availability}
\else
  \vspace{-0.1in}
  \section*{Code Availability}
\fi

The \textit{PSCMIA} framework source code and experimental artifacts are available at ~\href{https://anonymous.4open.science/r/PSCMIA}{https://anonymous.4open.science/r/PSCMIA}.
% We disclosed our findings, including the PoC code, to Broadcom and ARM in January 2025. ARM indicated that the issue appears to originate at the SoC/board level and recommended coordinated disclosure with Broadcom. However, at the time of writing, Broadcom has not responded.

\bibliographystyle{IEEEtran}
\bibliography{bibliography}

\vspace{-0.45in}

\begin{IEEEbiography}
[{\raisebox{0.30\height}{\includegraphics[width=1in,clip, keepaspectratio]{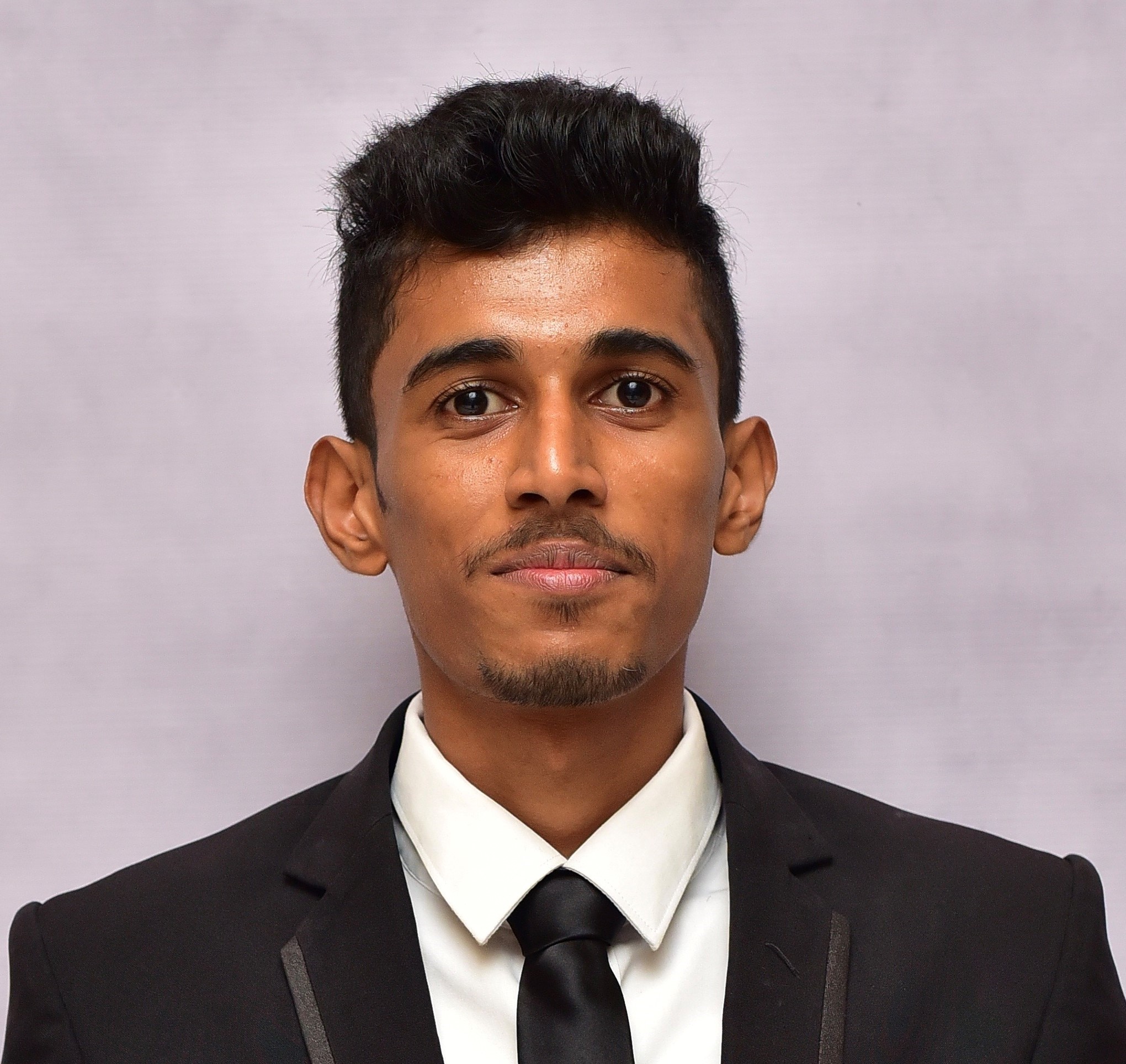}}}]{Sahan Sanjaya}
is a fourth-year Ph.D. student in the Department of Computer and Information Science and Engineering at the University of Florida. He received his B.Sc. degree in Electronic and Telecommunication Engineering from the University of Moratuwa, Sri Lanka, in 2022. His research interests include side-channel attacks, hardware security, pre-silicon validation, and post-silicon validation.
\end{IEEEbiography}

\vspace{-0.5in}

\begin{IEEEbiography}
[{\includegraphics[width=1in,height=1.25in,clip,keepaspectratio]{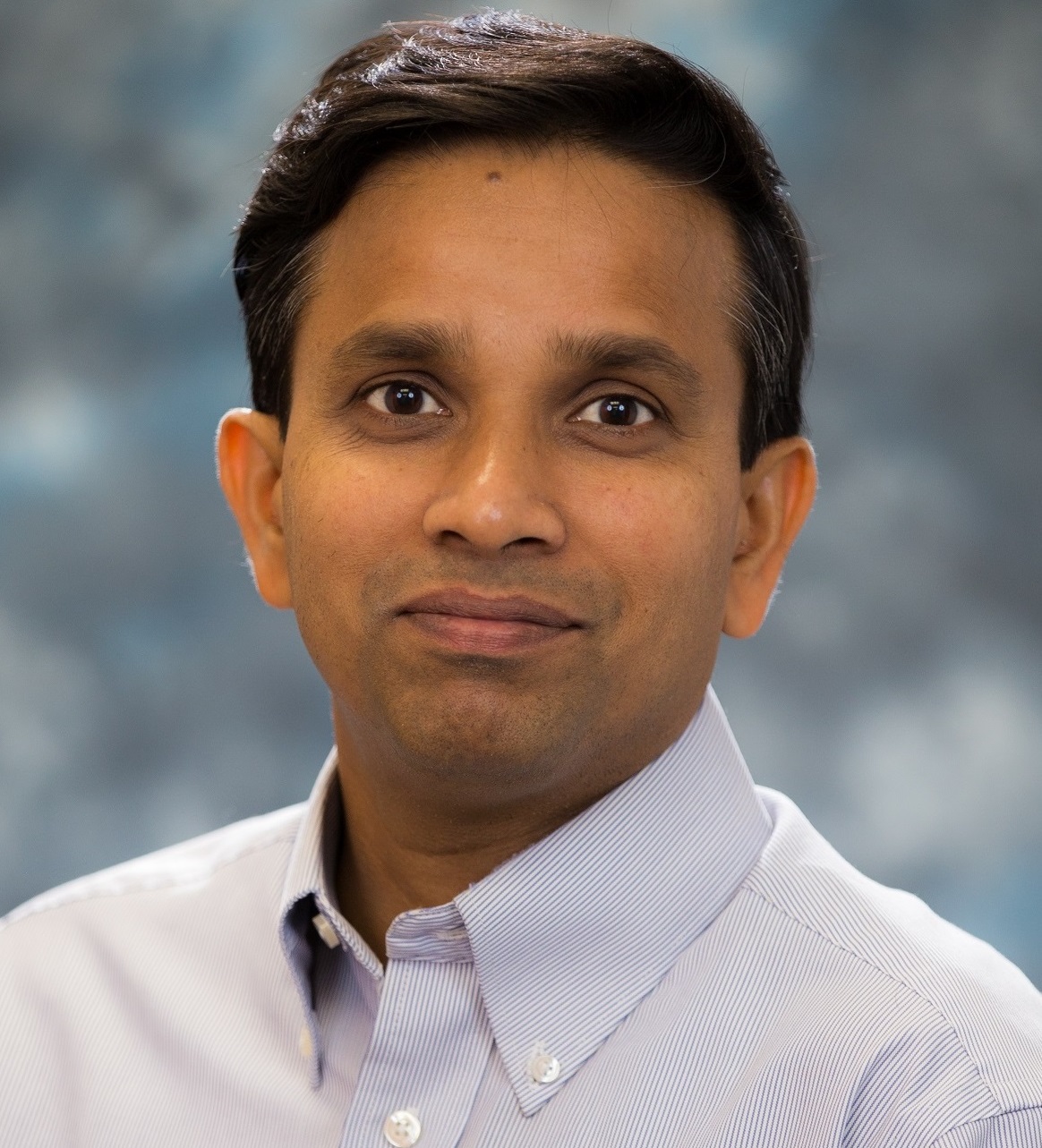}}]
{Prabhat Mishra}
is a Professor in the Department of Computer and Information Science and Engineering at the University of Florida. His research interests include embedded and cyber-physical systems, hardware security and trust, and energy-aware computing. He currently serves as an Associate Editor of ACM Transactions on Design Automation of Electronic Systems and ACM Transactions on Embedded Computing Systems. He is an IEEE Fellow, an AAAS Fellow, and an ACM Distinguished Scientist.
\end{IEEEbiography}

\newpage
\appendices
\section{PSCMIA Detailed Results}\label{sec:apx_per_class}

Tables~\ref{tab:fc_mia_results_full_sota} and~\ref{tab:cnn_mia_results_full_sota} compare PSCMIA against several representative output-based and label-only membership inference attacks across FC and CNN target models. The results show that the relative effectiveness of the attacks varies across datasets and model architectures. In particular, loss-based and shadow MIA attacks generally provide the strongest software-output baselines, while confidence-, entropy-, and correctness-based attacks are less consistent. PSCMIA remains competitive with these baselines and, in several configurations, achieves comparable or higher ROC-AUC while requiring only power side-channel observations rather than direct access to the target model's output probabilities.

\begin{table}[t]
\centering
\scriptsize
\setlength{\tabcolsep}{3.0pt}
\caption{Membership inference attack results comparison for FC target models. We compare probability vector-based shadow MIA, label-only MIA, confidence-, loss-, entropy-, and correctness-based attacks, and PSCMIA across datasets. PSCMIA is evaluated on STM32F3 and XMEGA.}
\label{tab:fc_mia_results_full_sota}
\begin{tabular}{llcccc}
\toprule
\textbf{Attack} &
\textbf{Dataset} &
\textbf{ROC-AUC} &
\textbf{FPR} &
\textbf{TPR@1\%FPR} &
\textbf{Acc.} \\
\midrule

\multirow{4}{*}{Shadow MIA~\cite{shokri2017membership}}
    & MNIST   & 0.6582 & 0.7090 & 0.0155 & 0.6138 \\
    & FMNIST  & 0.6804 & 0.7215 & 0.0265 & 0.6280 \\
    & CIFAR10 & 0.8794 & 0.1920 & 0.1635 & 0.8045 \\
    & CINIC10 & 0.9445 & 0.1365 & 0.3925 & 0.8782 \\
\midrule

\multirow{4}{*}{Label-only~\cite{li2025enhanced}}
    & MNIST   & 0.5281 & 0.3185 & 0.0305 & 0.5180 \\
    & FMNIST  & 0.6016 & 0.4715 & 0.0515 & 0.5773 \\
    & CIFAR10 & 0.8137 & 0.2355 & 0.0905 & 0.7720 \\
    & CINIC10 & 0.8960 & 0.2000 & 0.1730 & 0.8518 \\
\midrule

\multirow{4}{*}{Confidence~\cite{salem2018ml}}
    & MNIST   & 0.6408 & 0.6735 & 0.0115 & 0.6440 \\
    & FMNIST  & 0.6213 & 0.7275 & 0.0120 & 0.6255 \\
    & CIFAR10 & 0.7339 & 0.2665 & 0.0875 & 0.6770 \\
    & CINIC10 & 0.8157 & 0.2615 & 0.1360 & 0.7468 \\
\midrule

\multirow{4}{*}{Loss~\cite{yeom2018privacy}}
    & MNIST   & 0.6419 & 0.6685 & 0.0115 & 0.6460 \\
    & FMNIST  & 0.6317 & 0.7105 & 0.0135 & 0.6378 \\
    & CIFAR10 & 0.8698 & 0.2170 & 0.1450 & 0.7917 \\
    & CINIC10 & 0.9400 & 0.1455 & 0.2395 & 0.8735 \\
\midrule

\multirow{4}{*}{Entropy~\cite{salem2018ml}}
    & MNIST   & 0.6397 & 0.6815 & 0.0140 & 0.6418 \\
    & FMNIST  & 0.6208 & 0.7295 & 0.0135 & 0.6242 \\
    & CIFAR10 & 0.7085 & 0.2710 & 0.0665 & 0.6548 \\
    & CINIC10 & 0.8045 & 0.2785 & 0.1150 & 0.7260 \\
\midrule

\multirow{4}{*}{Correctness~\cite{yeom2018privacy}}
    & MNIST   & 0.5360 & 0.9280 & 0.0000 & 0.5360 \\
    & FMNIST  & 0.5862 & 0.8275 & 0.0000 & 0.5863 \\
    & CIFAR10 & 0.7920 & 0.2590 & 0.0000 & 0.7920 \\
    & CINIC10 & 0.8690 & 0.1950 & 0.0000 & 0.8690 \\
\midrule

\multirow{4}{*}{\shortstack[l]{PSCMIA\\(STM32F3)}}
    & MNIST   & 0.6193 & 0.1010 & 0.0140 & 0.5248 \\
    & FMNIST  & 0.6795 & 0.4340 & 0.0365 & 0.6238 \\
    & CIFAR10 & 0.7958 & 0.2540 & 0.0675 & 0.7165 \\
    & CINIC10 & 0.9065 & 0.1445 & 0.1965 & 0.8308 \\
\midrule

\multirow{4}{*}{\shortstack[l]{PSCMIA\\(XMEGA)}}
    & MNIST   & 0.6186 & 0.4885 & 0.0300 & 0.5810 \\
    & FMNIST  & 0.6477 & 0.5330 & 0.0145 & 0.6058 \\
    & CIFAR10 & 0.7952 & 0.2435 & 0.0885 & 0.7390 \\
    & CINIC10 & 0.8925 & 0.1760 & 0.2180 & 0.8328 \\
\bottomrule
\end{tabular}
\end{table}

\begin{table}[t]
\centering
\scriptsize
\setlength{\tabcolsep}{3.0pt}
\caption{Membership inference attack results comparison for CNN target models. We compare probability vector-based shadow MIA, label-only MIA, confidence-, loss-, entropy-, and correctness-based attacks, and PSCMIA across datasets. PSCMIA is evaluated on STM32F3 and XMEGA.}
\label{tab:cnn_mia_results_full_sota}
\begin{tabular}{llcccc}
\toprule
\textbf{Attack} &
\textbf{Dataset} &
\textbf{ROC-AUC} &
\textbf{FPR} &
\textbf{TPR@1\%FPR} &
\textbf{Acc.} \\
\midrule

\multirow{4}{*}{Shadow MIA~\cite{shokri2017membership}}
    & MNIST   & 0.6089 & 0.5040 & 0.0120 & 0.5690 \\
    & FMNIST  & 0.6945 & 0.6880 & 0.0300 & 0.6328 \\
    & CIFAR10 & 0.7726 & 0.3170 & 0.0780 & 0.7010 \\
    & CINIC10 & 0.8572 & 0.2290 & 0.0880 & 0.7755 \\
\midrule

\multirow{4}{*}{Label-only~\cite{li2025enhanced}}
    & MNIST   & 0.5111 & 0.1040 & 0.0235 & 0.5182 \\
    & FMNIST  & 0.5681 & 0.2300 & 0.0455 & 0.5725 \\
    & CIFAR10 & 0.7285 & 0.2420 & 0.0870 & 0.6895 \\
    & CINIC10 & 0.7978 & 0.2125 & 0.1315 & 0.7570 \\
\midrule

\multirow{4}{*}{Confidence~\cite{salem2018ml}}
    & MNIST   & 0.5935 & 0.5560 & 0.0110 & 0.5683 \\
    & FMNIST  & 0.6090 & 0.6650 & 0.0175 & 0.6270 \\
    & CIFAR10 & 0.6132 & 0.4015 & 0.0335 & 0.5853 \\
    & CINIC10 & 0.6620 & 0.4195 & 0.0435 & 0.6188 \\
\midrule

\multirow{4}{*}{Loss~\cite{yeom2018privacy}}
    & MNIST   & 0.5953 & 0.5545 & 0.0110 & 0.5690 \\
    & FMNIST  & 0.6217 & 0.6435 & 0.0175 & 0.6432 \\
    & CIFAR10 & 0.7661 & 0.3965 & 0.0355 & 0.7000 \\
    & CINIC10 & 0.8472 & 0.2790 & 0.0900 & 0.7768 \\
\midrule

\multirow{4}{*}{Entropy~\cite{salem2018ml}}
    & MNIST   & 0.5942 & 0.5420 & 0.0140 & 0.5637 \\
    & FMNIST  & 0.6079 & 0.6520 & 0.0200 & 0.6260 \\
    & CIFAR10 & 0.5976 & 0.4610 & 0.0235 & 0.5763 \\
    & CINIC10 & 0.6528 & 0.4020 & 0.0370 & 0.6162 \\
\midrule

\multirow{4}{*}{Correctness~\cite{yeom2018privacy}}
    & MNIST   & 0.5280 & 0.9415 & 0.0000 & 0.5280 \\
    & FMNIST  & 0.5867 & 0.8250 & 0.0000 & 0.5867 \\
    & CIFAR10 & 0.7035 & 0.3075 & 0.0000 & 0.7035 \\
    & CINIC10 & 0.7728 & 0.2430 & 0.0000 & 0.7728 \\
\midrule

\multirow{4}{*}{\shortstack[l]{PSCMIA\\(STM32F3)}}
    & MNIST   & 0.5821 & 0.0560 & 0.0140 & 0.5060 \\
    & FMNIST  & 0.6514 & 0.1370 & 0.0360 & 0.5545 \\
    & CIFAR10 & 0.7464 & 0.2360 & 0.0540 & 0.6495 \\
    & CINIC10 & 0.7417 & 0.2350 & 0.0405 & 0.6710 \\
\midrule

\multirow{4}{*}{\shortstack[l]{PSCMIA\\(XMEGA)}}
    & MNIST   & 0.5784 & 0.5165 & 0.0240 & 0.5448 \\
    & FMNIST  & 0.6588 & 0.4875 & 0.0215 & 0.6068 \\
    & CIFAR10 & 0.7668 & 0.3010 & 0.0050 & 0.7240 \\
    & CINIC10 & 0.7757 & 0.2995 & 0.0475 & 0.7160 \\
\bottomrule
\end{tabular}
\end{table}

Table~\ref{tab:app_pscmia_lr_et_rf} presents the PSCMIA results for the three evaluated attack models. The results show that PSCMIA achieves comparable performance across all three attack model types. However, ET slightly outperforms RF and LR in the majority of the evaluated configurations.

Table~\ref{tab:app_pscmia_et_perclass_auc} presents the per-class PSCMIA ROC-AUC results using ET across the evaluated datasets, target model architectures, and devices. The results show that membership leakage varies across target classes, with some classes exhibiting substantially stronger attack performance than others. Despite this class-wise variation, ET achieves above random ROC-AUC for nearly all classes, demonstrating that PSCMIA captures membership-dependent leakage across a broad range of class labels rather than relying on only a small subset of classes.

\begin{table}[h!]
\centering
\scriptsize
\setlength{\tabcolsep}{3.8pt}
\caption{PSCMIA performance of three representative attack classifiers: Logistic Regression (LR), Extra Trees (ET), and Random Forest (RF). All metrics are computed by pooling predictions across the ten target classes. Bold ROC-AUC indicates the best-performing classifier for each board-target-dataset configuration.}
\label{tab:app_pscmia_lr_et_rf}

\begin{tabular}{lllcccc}
\toprule
\textbf{Board} &
\textbf{Model/Dataset} &
\textbf{Attack} &
\textbf{ROC-AUC} &
\textbf{FPR} &
\textbf{Recall} &
\textbf{Acc.} \\
\midrule

STM32F3 & FC / MNIST
    & LR & 0.6113 & 0.1310 & 0.1925 & 0.5308 \\
        & 
    & ET & 0.6193 & 0.1010 & 0.1505 & 0.5248 \\
        &
    & RF & \textbf{0.6165} & 0.1045 & 0.1475 & 0.5215 \\
\addlinespace[2pt]

STM32F3 & FC / FMNIST
    & LR & 0.6708 & 0.4860 & 0.7290 & 0.6215 \\
        &
    & ET & \textbf{0.6795} & 0.4340 & 0.6815 & 0.6238 \\
        &
    & RF & 0.6736 & 0.3945 & 0.6125 & 0.6090 \\
\addlinespace[2pt]

STM32F3 & FC / CIFAR10
    & LR & 0.7771 & 0.3165 & 0.7190 & 0.7013 \\
        &
    & ET & \textbf{0.7958} & 0.2540 & 0.6870 & 0.7165 \\
        &
    & RF & 0.7893 & 0.2175 & 0.6345 & 0.7085 \\
\addlinespace[2pt]

STM32F3 & FC / CINIC10
    & LR & \textbf{0.9086} & 0.2270 & 0.8905 & 0.8318 \\
        &
    & ET & 0.9065 & 0.1445 & 0.8060 & 0.8308 \\
        &
    & RF & 0.9012 & 0.1605 & 0.7945 & 0.8170 \\

\midrule

STM32F3 & CNN / MNIST
    & LR & 0.5505 & 0.3915 & 0.4240 & 0.5163 \\
        &
    & ET & \textbf{0.5821} & 0.0560 & 0.0680 & 0.5060 \\
        &
    & RF & 0.5785 & 0.1095 & 0.1520 & 0.5213 \\
\addlinespace[2pt]

STM32F3 & CNN / FMNIST
    & LR & 0.6047 & 0.5580 & 0.7025 & 0.5723 \\
        &
    & ET & \textbf{0.6514} & 0.1370 & 0.2460 & 0.5545 \\
        &
    & RF & 0.6365 & 0.1545 & 0.2545 & 0.5500 \\
\addlinespace[2pt]

STM32F3 & CNN / CIFAR10
    & LR & 0.6928 & 0.3605 & 0.6365 & 0.6380 \\
        &
    & ET & \textbf{0.7464} & 0.2360 & 0.5350 & 0.6495 \\
        &
    & RF & 0.7371 & 0.2010 & 0.4690 & 0.6340 \\
\addlinespace[2pt]

STM32F3 & CNN / CINIC10
    & LR & \textbf{0.7519} & 0.4115 & 0.7885 & 0.6885 \\
        &
    & ET & 0.7417 & 0.2350 & 0.5770 & 0.6710 \\
        &
    & RF & 0.7273 & 0.2205 & 0.5295 & 0.6545 \\

\midrule

XMEGA & FC / MNIST
    & LR & 0.6037 & 0.5420 & 0.6770 & 0.5675 \\
      &
    & ET & 0.6186 & 0.4885 & 0.6505 & 0.5810 \\
      &
    & RF & \textbf{0.6216} & 0.4670 & 0.6320 & 0.5825 \\
\addlinespace[2pt]

XMEGA & FC / FMNIST
    & LR & \textbf{0.6587} & 0.5810 & 0.8065 & 0.6128 \\
      &
    & ET & 0.6477 & 0.5330 & 0.7445 & 0.6058 \\
      &
    & RF & 0.6526 & 0.5160 & 0.7410 & 0.6125 \\
\addlinespace[2pt]

XMEGA & FC / CIFAR10
    & LR & 0.7744 & 0.3235 & 0.7535 & 0.7150 \\
      &
    & ET & \textbf{0.7952} & 0.2435 & 0.7215 & 0.7390 \\
      &
    & RF & 0.7809 & 0.2375 & 0.7200 & 0.7413 \\
\addlinespace[2pt]

XMEGA & FC / CINIC10
    & LR & 0.8640 & 0.2245 & 0.8150 &  0.7953 \\
      &
    & ET & 0.8925 & 0.1760 & 0.8415 & 0.8328\\
      &
    & RF & \textbf{0.8927} & 0.1720 & 0.8460 & 0.8370\\

\midrule

XMEGA & CNN / MNIST
    & LR & 0.5727 & 0.5730 & 0.6590 & 0.5430 \\
      &
    & ET & \textbf{0.5784} & 0.5165 & 0.6060 & 0.5448 \\
      &
    & RF & 0.5767 & 0.4920 & 0.5930 & 0.5505 \\
\addlinespace[2pt]

XMEGA & CNN / FMNIST
    & LR & 0.6561 & 0.5170 & 0.7350 & 0.6090 \\
      &
    & ET & \textbf{0.6588} & 0.4875 & 0.7010 & 0.6067 \\
      &
    & RF & 0.6587 & 0.4735 & 0.6815 & 0.6040 \\
\addlinespace[2pt]

XMEGA & CNN / CIFAR10
    & LR & 0.7561 & 0.3545 & 0.7565 & 0.7010 \\
      &
    & ET & 0.7668 & 0.3010 & 0.7490 & 0.7240 \\
      &
    & RF & \textbf{0.7736} & 0.2965 & 0.7465 & 0.7250 \\
\addlinespace[2pt]

XMEGA & CNN / CINIC10
    & LR & 0.7627 & 0.3410 & 0.7660 & 0.7125 \\
      &
    & ET & 0.7757 & 0.2995 & 0.7315 & 0.7160 \\
      &
    & RF & \textbf{0.7825} & 0.2735 & 0.7415 & 0.7340 \\

\bottomrule
\end{tabular}
\end{table}

\section{Impact of Suppressing {argmax} Leakage}
\label{sec:apx_defense}

Here, we investigate whether suppressing the power leakage associated with the final argmax computation can mitigate PSCMIA. As discussed in Section~\ref{subsubsec:poi_identification}, the argmax-related trace region contains more membership-relevant information compared to other regions. Moreover, conventional MIAs benefit from access to detailed prediction outputs, motivating defenses that restrict such information. We therefore emulate a leakage-suppression countermeasure by removing the trace segment corresponding to the argmax computation before training and evaluating PSCMIA. This experiment evaluates whether eliminating leakage from this final prediction-related operation alone is sufficient to suppress membership inference.

Table~\ref{tab:pscmia_aginst_defense} shows that removing the argmax-related segment has only a limited effect on FC models. For example, the ROC-AUC decreases from $0.6795$ to $0.6661$ on FMNIST, from $0.7958$ to $0.7903$ on CIFAR10, and from $0.9065$ to $0.9060$ on CINIC10. In contrast, CNN models exhibit a larger reduction, particularly for CIFAR10 and CINIC10, where ROC-AUC decreases from $0.7464$ to $0.6742$ and from $0.7417$ to $0.6880$, respectively. Nevertheless, the attack remains above random guessing in all evaluated settings. These results indicate that membership leakage is not confined to the final argmax operation; earlier inference computations also retain membership-related information. Consequently, suppressing a single high-leakage region is insufficient to eliminate PSCMIA, suggesting that effective defenses should address leakage across multiple stages of the inference pipeline rather than only the final prediction computation.

\begin{table}[h!]
\centering
\scriptsize
\setlength{\tabcolsep}{2.8pt}
\caption{Impact of suppressing the argmax-related power-trace segment on PSCMIA. ``Without'' denotes the original attack without leakage suppression, while ``With'' denotes PSCMIA after removing the argmax-related trace segment.}
\label{tab:pscmia_aginst_defense}
\begin{tabular}{llccccc}
\toprule
\textbf{Defense} &
\textbf{Model} &
\textbf{ROC-AUC} &
\textbf{FPR} &
\textbf{Recall} &
\textbf{TPR@1\%FPR} &
\textbf{Acc.} \\
\midrule

\multirow{8}{*}{Without}
    & MNIST-FC   & 0.6193 & 0.1010 & 0.1505 & 0.0140  & 0.5248 \\
    & FMNIST-FC  & 0.6795 & 0.4340 & 0.6815 & 0.0365  & 0.6238 \\
    & CIFAR10-FC & 0.7958 & 0.2540 & 0.6870 & 0.0675  & 0.7165 \\
    & CINIC10-FC & 0.9065 & 0.1445 & 0.8060 & 0.1965 & 0.8308 \\
    & MNIST-CNN   & 0.5821 & 0.0560 & 0.0680 & 0.0140 & 0.5060 \\
    & FMNIST-CNN  & 0.6514 & 0.1370 & 0.2460 & 0.0360 & 0.5545 \\
    & CIFAR10-CNN & 0.7464 & 0.2360 & 0.5350 & 0.0540 & 0.6495 \\
    & CINIC10-CNN & 0.7417 & 0.2350 & 0.5770 & 0.0405 & 0.6710 \\
\midrule

\multirow{8}{*}{With}
    & MNIST-FC   & 0.6209 & 0.0635  & 0.1020  & 0.0160  & 0.5193 \\
    & FMNIST-FC  & 0.6661 & 0.2245  & 0.3790  & 0.0280  & 0.5773 \\
    & CIFAR10-FC & 0.7903 & 0.2210  & 0.6485  & 0.0600 & 0.7138\\
    & CINIC10-FC & 0.9060 & 0.1485  &  0.7890 & 0.1800 &  0.8203\\
    & MNIST-CNN   & 0.5783 &  0.0230 & 0.0275  & 0.0135   & 0.5023 \\
    & FMNIST-CNN  & 0.6238 & 0.1255 & 0.1905 & 0.0150 & 0.5325\\
    & CIFAR10-CNN & 0.6742 & 0.1600  &  0.3145  & 0.0370 & 0.5772 \\
    & CINIC10-CNN & 0.6880 & 0.2635 & 0.5390 & 0.0395 & 0.6378 \\
 
\bottomrule
\end{tabular}
\end{table}

\begin{table*}[t]
\centering
\scriptsize
\setlength{\tabcolsep}{3.0pt}
\caption{Per-class ROC-AUC of PSCMIA using the Extra Trees (ET) attack model across target models, datasets, and devices.}
\label{tab:app_pscmia_et_perclass_auc}
\resizebox{\textwidth}{!}{%
\begin{tabular}{lllcccccccccc}
\toprule
\textbf{Board} &
\textbf{Model} &
\textbf{Dataset} &
\textbf{Class-0} &
\textbf{Class-1} &
\textbf{Class-2} &
\textbf{Class-3} &
\textbf{Class-4} &
\textbf{Class-5} &
\textbf{Class-6} &
\textbf{Class-7} &
\textbf{Class-8} &
\textbf{Class-9} \\
\midrule

STM32F3 & FC & MNIST
& 0.5776 & 0.5682 & 0.5877 & 0.7035 & 0.6838
& 0.6132 & 0.5749 & 0.6649 & 0.6832 & 0.6624 \\

STM32F3 & FC & FMNIST
& 0.7140 & 0.5722 & 0.6939 & 0.5898 & 0.7181
& 0.6187 & 0.8078 & 0.6644 & 0.5871 & 0.5979 \\

STM32F3 & FC & CIFAR10
& 0.7631 & 0.8280 & 0.8538 & 0.8242 & 0.8500
& 0.8205 & 0.8307 & 0.8039 & 0.8031 & 0.7381 \\

STM32F3 & FC & CINIC10
& 0.8729 & 0.8940 & 0.9422 & 0.8915 & 0.9092
& 0.9079 & 0.9244 & 0.9083 & 0.8992 & 0.9011 \\

\midrule

STM32F3 & CNN & MNIST
& 0.5373 & 0.6042 & 0.6351 & 0.6460 & 0.6080
& 0.5289 & 0.5463 & 0.5694 & 0.6241 & 0.6348 \\

STM32F3 & CNN & FMNIST
& 0.6965 & 0.5592 & 0.7555 & 0.6349 & 0.7527
& 0.5943 & 0.8637 & 0.5817 & 0.5631 & 0.5442 \\

STM32F3 & CNN & CIFAR10
& 0.7113 & 0.8269 & 0.8055 & 0.7854 & 0.8329
& 0.7650 & 0.8067 & 0.7270 & 0.6955 & 0.7507 \\

STM32F3 & CNN & CINIC10
& 0.7280 & 0.8237 & 0.8224 & 0.7551 & 0.8182
& 0.7949 & 0.8125 & 0.8039 & 0.7743 & 0.7068 \\

\midrule

XMEGA & FC & MNIST
& 0.5218 & 0.5535 & 0.6292 & 0.6594 & 0.6016
& 0.6371 & 0.6284 & 0.6091 & 0.6990 & 0.6151 \\

XMEGA & FC & FMNIST
& 0.6671 & 0.6095 & 0.6811 & 0.5950 & 0.7375
& 0.5688 & 0.7889 & 0.5507 & 0.5474 & 0.5148 \\

XMEGA & FC & CIFAR10
& 0.7744 & 0.8507 & 0.8475 & 0.8080 & 0.8056
& 0.8072 & 0.7711 & 0.7331 & 0.7410 & 0.7730 \\

XMEGA & FC & CINIC10
& 0.8235 & 0.9224 & 0.9413 & 0.8803 & 0.8951
& 0.8987 & 0.9018 & 0.8585 & 0.8899 & 0.8758 \\

\midrule

XMEGA & CNN & MNIST
& 0.5452 & 0.5479 & 0.5586 & 0.6124 & 0.6157
& 0.4714 & 0.5794 & 0.5325 & 0.6115 & 0.6531 \\

XMEGA & CNN & FMNIST
& 0.6670 & 0.5844 & 0.7304 & 0.6501 & 0.7248
& 0.5703 & 0.7893 & 0.5706 & 0.5682 & 0.4939 \\

XMEGA & CNN & CIFAR10
& 0.6994 & 0.8443 & 0.8263 & 0.7734 & 0.8307
& 0.7614 & 0.7712 & 0.7209 & 0.6640 & 0.7182 \\

XMEGA & CNN & CINIC10
& 0.7303 & 0.8024 & 0.8562 & 0.7590 & 0.7741
& 0.7409 & 0.8133 & 0.7875 & 0.7590 & 0.7268 \\

\bottomrule
\end{tabular}%
}
\end{table*}

% \section{Raw Traces vs Selected Features}

% \section{Impact of the Number of Shadow Models}

% \begin{figure}[t]
% \centering
% \begin{subfigure}{\columnwidth}
% \centering
% \includegraphics[width=\textwidth]{fig/CIFAR10_FCMNIST_pro_shadow_count_roc_auc_styles_per_class_models_logreg-extra_trees-rf.pdf}
% \caption{ROC-AUC.}
% \label{fig:shadow_models_auc_cifar10_pro}
% \end{subfigure}
% %
% \begin{subfigure}{\columnwidth}
% \centering
% \includegraphics[width=\textwidth]{fig/CIFAR10_FCMNIST_pro_shadow_count_accuracy_styles_per_class_models_logreg-extra_trees-rf.pdf}
% \caption{Attack accuracy.}
% \label{fig:shadow_models_acc_cifar10_pro}
% \end{subfigure}
% %
% \begin{subfigure}{\columnwidth}
% \centering
% \includegraphics[width=\textwidth]{fig/CIFAR10_FCMNIST_pro_shadow_count_fpr_percent_styles_per_class_models_logreg-extra_trees-rf.pdf}
% \caption{FPR.}
% \label{fig:shadow_models_fpr_cifar10_pro}
% \end{subfigure}
% \caption{Impact of the number of shadow datasets on membership inference attack performance for an FC target model trained on CIFAR10 and CINIC10 and deployed on STM32F3.}
% \label{fig:shadow_models_impact}
% \end{figure}

\section{Robustness to Noisy Power Traces}
\label{sec:apx_noise}

In practical attack settings, power-trace measurements are subject to measurement noise arising from the activity of other components on the device, environmental interference, acquisition equipment, and run-to-run measurement variability. To evaluate the robustness of PSCMIA under noisy conditions, we synthetically inject zero-mean Gaussian noise into each averaged power trace immediately after it is loaded from the trace dataset and before feature extraction. Given an original trace $\mathbf{x}_i$, the corresponding noisy trace is generated as
\[
\tilde{\mathbf{x}}_i=\mathbf{x}_i+\epsilon\sigma_i\mathbf{z}_i,
\]
where $\mathbf{z}_i\sim\mathcal{N}(\mathbf{0},\mathbf{I})$, $\sigma_i$ denotes the standard deviation of the original trace, and $\epsilon$ controls the relative noise magnitude. This trace-dependent scaling makes $\epsilon$ dimensionless and ensures that the injected noise is proportional to the signal variation of each measurement rather than to its absolute amplitude. We vary $\epsilon$ from the noiseless setting, $\epsilon=0$, to increasingly severe noise levels and rerun the complete attack pipeline at each setting using noisy profiling and evaluation traces. We then compare ROC-AUC and accuracy across noise levels.

\begin{figure}[h!]
\centering
\begin{subfigure}{\columnwidth}
    \centering
    \includegraphics[width=\textwidth]{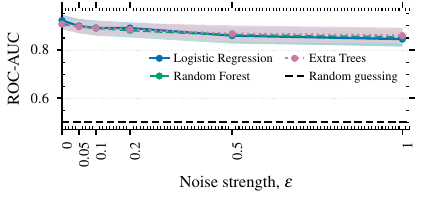}
    \caption{ROC-AUC under increasing Gaussian noise.\vspace{0.2in}}
    \label{fig:noise_robustness_auc}
\end{subfigure}

\begin{subfigure}{\columnwidth}
    \centering
    \includegraphics[width=\textwidth]{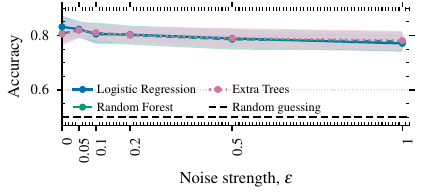}
    \caption{Attack accuracy under increasing Gaussian noise.}
    \label{fig:noise_robustness_accuracy}
\end{subfigure}
\caption{Robustness of PSCMIA to trace noise. Evaluated on CINIC10-FC on STM32F3}
\label{fig:noise_robustness}
\end{figure}

Figure~\ref{fig:noise_robustness} evaluates the robustness of PSCMIA when zero-mean Gaussian noise is added to the averaged power traces. At $\epsilon=0$, the evaluated attack models trained on the CINIC10 dataset achieve the highest ROC-AUC. Increasing the noise strength causes a gradual, rather than abrupt, reduction in attack effectiveness. At the strongest evaluated noise level, $\epsilon=1$, where the noise and power trace standard deviations are approximately equal, ROC-AUC remains approximately $0.85$-$0.86$, while accuracy remains approximately $0.78$, both substantially above the random-guessing baseline of $0.5$. The similar degradation trends across classifier families indicate that the observed robustness is not specific to one attack-model architecture. The shaded regions indicate $\pm1$ standard deviation across classes and become moderately wider as $\epsilon$ increases, indicating that some target classes are more sensitive to measurement degradation than others. Nevertheless, the consistently high performance across the entire noise range demonstrates that PSCMIA does not rely exclusively on small, noise-sensitive trace variations and remains effective under considerable corruption of the measured power trace.

\section{Attack Model Transferability Results}
\label{sec:apx_transferability}

\begin{figure*}[t]
    \centering

    % Row 1
    \begin{subfigure}[t]{0.48\textwidth}
        \centering
        \includegraphics[width=\linewidth]{fig/extra_trees_roc_auc_transfer_heatmap_STM_FC.pdf}
        \caption{FC - STM32F3}
        \label{fig:transfer_fc_stm32f3}
    \end{subfigure}
    \hfill
    \begin{subfigure}[t]{0.48\textwidth}
        \centering
        \includegraphics[width=\linewidth]{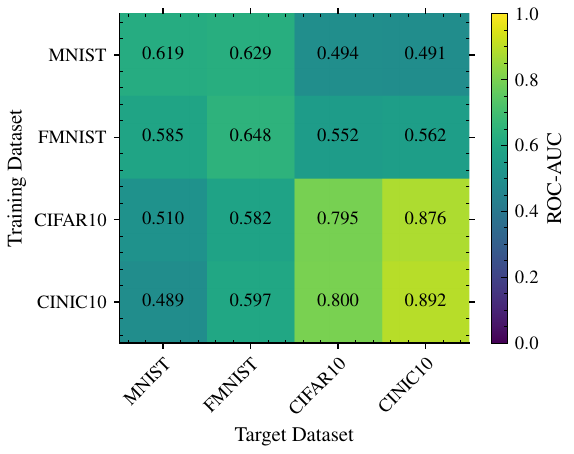}
        \caption{FC - XMEGA}
        \label{fig:transfer_fc_xmega}
    \end{subfigure}

    % Row 2
    \begin{subfigure}[t]{0.48\textwidth}
        \centering
        \includegraphics[width=\linewidth]{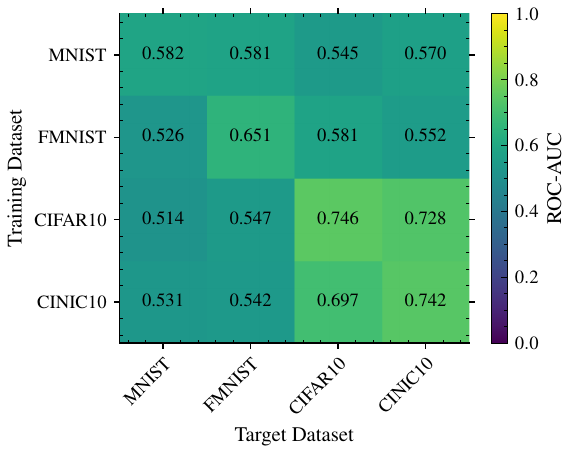}
        \caption{CNN - STM32F3}
        \label{fig:transfer_cnn_stm32f3}
    \end{subfigure}
    \hfill
    \begin{subfigure}[t]{0.48\textwidth}
        \centering
        \includegraphics[width=\linewidth]{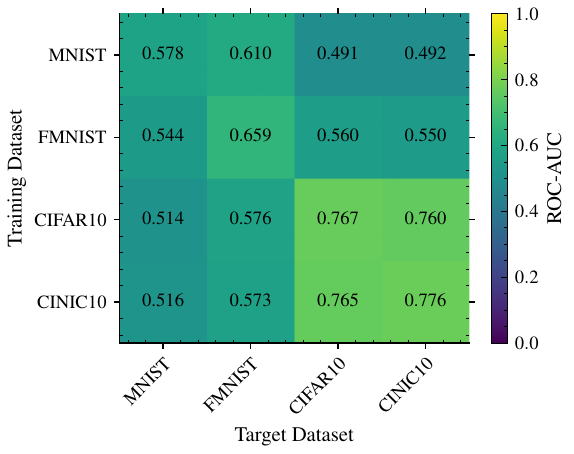}
        \caption{CNN - XMEGA}
        \label{fig:transfer_cnn_xmega}
    \end{subfigure}

    \caption{Transferability of PSCMIA across datasets and target-model architectures on STM32F3 and XMEGA. Rows within each heatmap indicate the source dataset used to train the attack model, while columns indicate the target dataset used for evaluation.}
    \label{fig:apx_attack_transferability}
\end{figure*}

Figure~\ref{fig:apx_attack_transferability} presents the cross-dataset transferability of PSCMIA for FC and CNN models on both STM32F3 and XMEGA. The diagonal entries represent same-dataset attack performance, while the off-diagonal ROC-AUC values quantify how well an attack model trained on one dataset transfers to another, highlighting the dataset dependence of the learned side-channel leakage patterns.

\end{document}